\documentclass[
    sigconf,
    review=false,
    authordraft=false,
    natbib=true,
    nonacm=true,
]{acmart}
\AtBeginDocument{%
  }

\setcopyright{rightsretained}
\copyrightyear{2024}
\acmYear{2024}
\acmDOI{XXXXXXX.XXXXXXX}

\usepackage{algorithmic}
\usepackage{graphicx}
\usepackage{textcomp}
\usepackage{booktabs} 
\usepackage{multirow}
\usepackage{bigdelim} 
\usepackage{tabularx}
\usepackage{centernot}
\usepackage{csquotes}
\usepackage{comment}

\usepackage{colortbl} 
\usepackage{array}
\definecolor{dkgreen}{rgb}{0,0.6,0}
\definecolor{gray}{rgb}{0.5,0.5,0.5}
\definecolor{mauve}{rgb}{0.58,0,0.82}
\definecolor{linkcolor}{rgb}{0.65,0,0}
\definecolor{citecolor}{rgb}{0,0.4,0}
\definecolor{urlcolor}{rgb}{0,0,0.65}
\definecolor{nicegreenfill}{RGB}{213,232,212}
\definecolor{nicegreenborder}{RGB}{151,192,128}
\definecolor{niceorangefill}{RGB}{255,230,204}
\definecolor{niceorangeborder}{RGB}{225,178,61}
\definecolor{niceyellowfill}{RGB}{255,242,204}
\definecolor{niceyellowborder}{RGB}{222,195,114}
\definecolor{niceredfill}{RGB}{171,70,70}
\definecolor{niceredborder}{RGB}{150,38,38}
\definecolor{nicebluefill}{RGB}{218,232,252}
\definecolor{niceblueborder}{RGB}{133,161,203}
\definecolor{plotblue}{rgb}{0.00000,0.44700,0.74100}
\definecolor{plotorange}{rgb}{0.85000,0.32500,0.09800}

\usepackage{listings}
\usepackage{tikz} 
\usetikzlibrary{external,calc,shapes.geometric,arrows}
\usepackage{tikzpeople} 

\usepackage{pifont}
\newcommand{\cmark}{\ding{51}}%
\newcommand{\xmark}{\ding{55}}%
\newcommand{\pmark}{\ding{58}}%

\usepackage{fontawesome} 
\usepackage[nolist]{acronym}

\usepackage{hyperref}
\hypersetup{
    hidelinks,
    colorlinks=true
}

\usepackage[
    colorinlistoftodos,
    prependcaption,
    textwidth=\marginparwidth,
    textsize=small
]{todonotes} 

\newcommand{\ruggedtodo}[2][]{\tikzexternaldisable\todo[#1]{#2}\tikzexternalenable}
\ifdefined\COMMENTSOFF
  \renewcommand{\ruggedtodo}[2][]{}
\fi

\newcommand{\repmov}{{\tt REP MOV}}
\newcommand{\novuln}{\textcolor{gray}{N/A}}
\newcommand{\noprotect}{\textcolor{niceredborder}{\xmark}}
\newcommand{\protects}{\textcolor{nicegreenborder}{\cmark}}
\newcommand{\protectswith}{\textcolor{niceyellowborder}{\pmark}}
\newcommand{\protectskey}{
    \protects{} -- mitigation prevents side-channel attack. \par
    \protectswith{} -- mitigation prevents attack only in combination with other mitigation(s) marked with \protectswith.\par
    \noprotect{} -- mitigation has no effect on this attack.
}
\newcommand{\channel}{{$\Longrightarrow$}}
\newcommand{\nochannel}{{$\centernot\Longrightarrow$}}
\newcommand{\channelkey}{
\channel{} -- Side-channel leakage is observable with all mitigations disabled. \par
\nochannel{} -- Side-channel leakage is not observable.
}
\newcommand{\pinned}{\faThumbTack}

\title{Microarchitectural Security of AWS Firecracker VMM for Serverless Cloud Platforms}

\author{Zane Weissman}
\email{zweissman@wpi.edu}
\orcid{0000-0002-8594-5390}
\affiliation{
    \institution{Worcester Polytechnic Institute}
    \streetaddress{100 Institute Road}
    \city{Worcester}
    \state{MA}
    \country{USA}
    \postcode{01609-2280}
}

\author{Thore Tiemann}
\email{t.tiemann@uni-luebeck.de}
\orcid{0000-0001-9018-4226}
\affiliation{
    \institution{University of L{\"u}beck}
    \streetaddress{Ratzeburger Allee 160}
    \city{Lübeck}
    \state{S-H}
    \country{Germany}
    \postcode{23562}
}

\author{Thomas Eisenbarth}
\email{thomas.eisenbarth@uni-luebeck.de}
\orcid{0000-0003-1116-6973}
\affiliation{
    \institution{University of L{\"u}beck}
    \streetaddress{Ratzeburger Allee 160}
    \city{Lübeck}
    \state{S-H}
    \country{Germany}
    \postcode{23562}
}

\author{Berk Sunar}
\email{sunar@wpi.edu}
\orcid{0000-0001-5404-5368}
\affiliation{
    \institution{Worcester Polytechnic Institute}
    \streetaddress{100 Institute Road}
    \city{Worcester}
    \state{MA}
    \country{USA}
    \postcode{01609-2280}
}

\begin{document}

\begin{acronym}[RIDL]
    \acro{AWS}[AWS]{Amazon Web Services}
    \acro{FaaS}[FaaS]{function-as-a-service}
    \acro{CaaS}[CaaS]{container-as-a-service}
    \acro{BPU}[BPU]{branch prediction unit}
    \acro{CSP}[CSP]{cloud service provider}
    \acro{KVM}[KVM]{kernel-based virtual machine}
    \acro{LFB}[LFB]{line-fill buffer}
    \acro{MDS}[MDS]{microarchitectural data sampling}
    \acro{OS}[OS]{operating system}
    \acro{PoC}[PoC]{proof-of-concept}
    \acro{SMT}[SMT]{simultaneous multi-threading}
    \acro{TAA}[TAA]{TSX Asynchronous Abort}
    \acro{VM}[VM]{virtual machine}
    \acro{VMM}[VMM]{virtual machine manager}
    \acro{RIDL}[RIDL]{Rogue In-flight Data Load}
\end{acronym}

\begin{abstract}
Firecracker is a virtual machine manager (VMM) purpose-built by Amazon Web Services (AWS) for serverless cloud platforms---services that run code for end users on a per-task basis, automatically managing server infrastructure.
Firecracker provides fast and lightweight VMs and promises a combination of the speed of containers, typically used to isolate small tasks, and the security of VMs, which tend to provide greater isolation at the cost of performance.
This combination of security and efficiency, AWS claims, makes it not only possible but safe to run thousands of user tasks from different users on the same hardware, with the host system rapidly and frequently switching between active tasks.
Though AWS states that microarchitectural attacks are included in their threat model, this class of attacks directly relies on shared hardware, just as the scalability of serverless computing relies on sharing hardware between unprecedented numbers of users.

In this work, we investigate just how secure Firecracker is against microarchitectural attacks.
First, we review Firecracker's stated isolation model and recommended best practices for deployment, identify potential threat models for serverless platforms, and analyze potential weak points. 
Then, we use microarchitectural attack proof-of-concepts to test the isolation provided by Firecracker and find that it offers little protection against Spectre or MDS attacks.
We discover two particularly concerning cases: \textbf{1)} a Medusa variant that threatens Firecracker VMs but \emph{not} processes running outside them, and is not mitigated by defenses recommended by AWS, and \textbf{2)} a Spectre-PHT variant that remains exploitable even if recommended countermeasures are in place and SMT is disabled in the system.
In summary, we show that AWS overstates the security inherent to the Firecracker VMM and provides incomplete guidance for properly securing cloud systems that use Firecracker.
\end{abstract}




\maketitle

\section{Introduction}
Serverless computing is an emerging trend in cloud computing where \acp{CSP} serve runtime environments to their customers.
This way, customers can focus on maintaining their function code while leaving the administrative work related to hardware, \ac{OS}, and sometimes runtime to the \acp{CSP}.
Common serverless platform models include \ac{FaaS} and \ac{CaaS}.
Since individual functions are typically small, but customers' applications can each be running anywhere from one to thousands of functions, \acp{CSP} aim for fitting as many functions on a single server as possible to minimize idle times and, in turn, maximize profit.
A rather light-weight approach to serving runtime environments is to run containers, which encapsulate a process with its dependencies so that only the necessary files for each process are loaded in virtual filesystems top of a shared kernel.
This reduces a switch between containers to little more than a context switch between processes.
On the other hand, full virtualization provides good isolation between \acp{VM} and therefore security between tenants, while being rather heavy-weight as each \ac{VM} comes with its own kernel.

Neither of these approaches, container or \ac{VM}, is ideal for use in serverless environments, where ideally many short-lived functions owned by many users will run simultaneously and switch often, so new mechanisms of isolation have been developed for this use case.
For example, mechanisms for in-process isolation~\cite{schrammel2020donky,vahldiek2019erim,narayan2021swivel} set out to improve the security of containers by reducing the attack surface of the runtime and underlying kernel. 
Protecting the kernel is important, as a compromised kernel directly leads to a fully compromised system in the container case. 
However, certain powerful protections, like limiting syscalls, also limit the functionality that is available to the container and even break compatibility with some applications.
In \ac{VM} research, developers created ever smaller and more efficient \acp{VM}, eventually leading to so-called microVMs.
MicroVMs provide the same isolation guarantees as usual virtual machines, but are very limited in their capabilities when it comes to device or \ac{OS} support, which makes them more light-weight compared to usual \acp{VM} and therefore better suited for serverless computing.

Firecracker~\cite{agache2020firecracker} is a \ac{VMM} designed to run microVMs while providing memory overhead and start times comparable to those of common container systems.
Firecracker is actively developed by \ac{AWS} and has been used in production for \acs{AWS} Lambda~\cite{amazon2023lambda} and \acs{AWS} Fargate~\cite{amazon2023fargate} serverless compute services since 2018~\cite{agache2020firecracker}.
\ac{AWS}'s design paper~\cite{agache2020firecracker} describes the features of Firecracker, how it diverges from more traditional virtual machines, and the intended isolation model that it provides: safety for \emph{``multiple functions run[ning] on the same hardware, protected against privilege escalation, information disclosure, covert channels, and other risks''}~\cite{agache2020firecracker}.
Furthermore, \ac{AWS} provides production host setup recommendations~\cite{amazon2023prodhostsetup} for securing parts of the CPU and kernel that a Firecracker \ac{VM} interacts with.
\emph{In this paper, we challenge the claim that Firecracker protects functions from covert and side-channels across microVMs.
We show that Firecracker itself does not add to the microarchitectural attack countermeasures but fully relies on the host and guest Linux kernels and CPU firmware/microcode updates.}

Microarchitectural attacks like the various Spectre~\cite{kocher2019spectre,koruyeh2018spectreReturns,canela2019systematicevaluation,horn2018sectrev4,maisuradze2018ret2spec,barberis2022branchhistoryinjection,wikner2022retbleed} and \ac{MDS}~\cite{vanschaik2019ridl,canella2019fallout,schwarz2019zombieload,moghimi2020medusa} variants pose a threat to multi-tenant systems as they are often able to bypass both software and architectural isolation boundaries, including those of \acp{VM}.
Spectre and \ac{MDS} threaten tenants that share CPU core resources like the \ac{BPU} or the \ac{LFB}.
\acp{CSP} providing more traditional services can mitigate the problem of shared hardware resources by pinning the long-lived \acp{VM} tenants to separate CPU cores, which effectively partitions the resources between the tenants and ensures that the microarchitectural state is only effected by a single tenant at a time.

In serverless environments, however, the threat of microarchitectural attacks is greater.
The reason for this is the short-livedness of the functions that are run by the different tenants.
Server resources in serverless environments are expected to be over-committed, which leads to tenant functions competing for compute resources on the same hardware.
Disabling \ac{SMT}, which would disable the concurrent use of CPU resources by two sibling threads, reduces the compute power of a CPU by up to 30\%~\cite{intel2002hyperthreading}.
If customers rent specific CPU cores, this performance penalty may be acceptable, or both threads on a CPU core might be rented together.
But for serverless services, the performance penalty directly translates to 30\% fewer customers that can be served in a given amount of time.
This is why it has to be assumed that most serverless \acp{CSP} keep \ac{SMT} enabled in their systems unless they state otherwise. 
The microarchitectural attack surface is largest if \ac{SMT} is enabled and the malicious thread has concurrent access to a shared core.
But there are also attack variants that perform just as well if the attacker thread prepares the microarchitecture before it yields the CPU core to the victim thread or executes right after the victim thread has paused execution.
And even if \ac{SMT} is disabled by the \ac{CSP} (as is the case for \acs{AWS} Lambda), tenants still share CPUs with multiple others in this time-sliced fashion.

\ac{AWS} claims that Firecracker running on a system with up-to-date microarchitectural defenses will provide sufficient hardening against microarchitectural attacks~\cite{agache2020firecracker}.
The Firecracker documentation also contains specific recommendations for microarchitectural security measures that should be enabled.
\emph{In this work, we examine Firecracker's security claims and recommendations and reveal oversights in its guidance as well as wholly unmitigated threats.}


In summary, our main contributions are:
\begin{itemize}
    \item We provide a comprehensive security analysis of the cross-tenant and tenant-hypervisor isolation of serverless compute when based on Firecracker \ac{VM}.
    \item We test Firecracker's defense capabilities against microarchitectural attack \acp{PoC}, employing protections available through microcode updates and the Linux kernel. We show that the virtual machine itself provides \emph{negligible protection} against major classes of microarchitectural attacks.
    \item We identify a variant of the Medusa \ac{MDS} attack that becomes exploitable from within Firecracker \acp{VM} \emph{even though it is not present on the host}. The kernel mitigation that protects against this exploit, and most known Medusa variants, is not mentioned by \ac{AWS}'s Firecracker host setup recommendations. Additionally, we show that disabling \ac{SMT} provides insufficient protection against the identified Medusa variant which urges the need of the kernel mitigation.
    \item We identify Spectre-PHT and Spectre-BTB variants which leak data even with recommended countermeasures in place. The Spectre-PHT variants even remain a problem when \ac{SMT} is disabled if the attacker and victim share a CPU core in a time-sliced fashion.
\end{itemize}

\subsection{Responsible Disclosure}
We informed the AWS security team about our findings and discussed technical details.
The AWS security team claims that the AWS services are not affected by our findings due to additional security measurements.
AWS agreed that Firecracker does not provide micro-architectural security on its own but only in combination with microcode updates and secure host and guest operating systems and plans to update its host setup recommendations for Firecracker installations.


\section{Background}


\subsection{\acs{KVM}}



The Linux \ac{KVM}~\cite{kivity2007kvm} provides an abstraction of the hardware-assisted virtualization features like Intel VT-x or AMD-V that are available in modern CPUs.
To support near-native execution, a \emph{guest mode} is added to the Linux kernel in addition to the existing \emph{kernel mode} and \emph{user mode}. 
If in Linux guest mode, \ac{KVM} causes the hardware to enter the hardware virtualization mode which replicates ring 0 and ring 3 privileges.\footnote{The virtualized ring 0 and ring 3 are one of the core reasons why near-native code execution is achieved.}

With \ac{KVM}, I/O virtualization is performed mostly in user space by the process that created the \ac{VM}, referred to as the \ac{VMM} or hypervisor, in contrast to earlier hypervisors which typically required a separate hypervisor process~\cite{qumranet2006kvm}.
A \ac{KVM} hypervisor provides each \ac{VM} guest with its own memory region that is separate from the memory region of the process that created the guest. 
This is true for guests created from kernel space as well as from user space. 
Each \ac{VM} maps to a process on the Linux host and each virtual CPU assigned to the guest is a thread in that host process.
The \ac{VM}'s userspace hypervisor process makes system calls to \ac{KVM} only when privileged execution is required, minimizing context switching and reducing the \ac{VM} to kernel attack surface.
Besides driving performance improvements across all sorts of applications, this design has allowed for the development of lightweight hypervisors that are especially useful for sandboxing individual programs and supporting cloud environments where many \acp{VM} are running at the same time.


\subsection{Serverless Cloud Computing}
An increasingly popular model for cloud computing is serverless computing, in which the \ac{CSP} manages scalability and availability of the servers that run the user's code.
One implementation of serverless computing is called \acf{FaaS}.
In this model, a cloud user defines functions that are called as necessary through the service provider's application programming interface (API) (hence the name ``\acl{FaaS}'') and the \ac{CSP} manages resource allocation on the server that executes the user's function (hence the name ``serverless computing''---the user does no server management).
Similarly, \acf{CaaS} computing runs containers, portable runtime packages, on demand.
The centralized server management of \ac{FaaS} and \ac{CaaS} is economically attractive to both \acp{CSP} and users.
The \ac{CSP} can manage its users' workloads however it pleases, optimize for minimal operating cost, and implement flexible pricing where users pay for the execution time that they use.
The user does not need to worry about server infrastructure design or management, and so reduces development costs and outsources maintenance cost to the \ac{CSP} at a relatively small and predictable rate.

\subsection{MicroVMs and \acs{AWS} Firecracker}
\ac{FaaS} and \ac{CaaS} providers use a variety of systems to manage running functions and containers.
Container systems like Docker, Podman, and LXD provide a convenient and lightweight way to package and run sandboxed applications in any environment.
However, compared to the virtual machines used for many more traditional forms of cloud computing, containers offer less isolation and therefore less security.
In recent years, major \acp{CSP} have introduced microVMs that back traditional containers with lightweight virtualization for extra security.~\cite{agache2020firecracker,young2019true}
The efficiency of hardware virtualization with \ac{KVM} and lightweight design of microVMs means that code in virtualized, containerized or container-like systems can run nearly as fast as unvirtualized code and with comparable overhead to a traditional container.

Firecracker~\cite{agache2020firecracker} is a microVM developed by \ac{AWS} to isolate each of the \acs{AWS} Lambda \ac{FaaS} and \acs{AWS} Fargate \ac{CaaS} workloads in a separate \ac{VM}. 
It only supports Linux guests on x86 or ARM Linux-KVM hosts and provides a limited number of devices that are available to guest systems.
These limitations allow Firecracker to be very light-weight in the size of its code base and in memory overhead for a running \ac{VM}, as well as very quick to boot or shut down.
Additionally, the use of \ac{KVM} lightens the requirements of Firecracker, since some virtualization functions are handled by kernel system calls and the host \ac{OS} manages \acp{VM} as standard processes.
Because of its small code base written in Rust, Firecracker is assumed to be very secure, even though security flaws have been identified in earlier versions (see \href{https://www.cve.org/CVERecord?id=CVE-2019-18960}{CVE-2019-18960}). 
Interestingly, the Firecracker white paper declares microarchitectural attacks to be in-scope of its attacker model~\cite{agache2020firecracker} but lacks a detailed security analysis or special countermeasures against microarchitectural attacks beyond common secure system configuration recommendations for the guest and host kernel.
The Firecracker documentation does provide system security recommendations~\cite{amazon2023prodhostsetup} that include a specific list of countermeasures, which we cover in \autoref{sec:firecracker-security-recs}.

\subsection{Meltdown and \acs{MDS}}
\label{mds-background}
In 2018, the Meltdown~\cite{lipp2018meltdown} attack showed that speculatively accessed data could be exfiltrated across security boundaries by encoding it into a cache side-channel.
This soon led to a whole class of similar attacks, known as \acf{MDS}, including Fallout~\cite{canella2019fallout}, \ac{RIDL}~\cite{vanschaik2019ridl}, \ac{TAA}~\cite{vanschaik2019ridl}, and Zombieload~\cite{schwarz2019zombieload}.
These attacks all follow the same general pattern to exploit speculative execution:
\begin{enumerate}
    \item The victim runs a program that handles secret data, and the secret data passes through a cache or CPU buffer.
    \item The attacker runs a specifically chosen instruction that will cause the CPU to mistakenly predict that the secret data will be needed. The CPU forwards the secret data to the attacker's instruction.
    \item The forwarded secret data is used as the index for a memory read to an array that the attacker is authorized to access, causing a particular line of that array to be cached.
    \item The CPU finishes checking the data and decides that the secret data was forwarded incorrectly, and reverts the execution state to before it was forwarded, but the state of the cache is not reverted. 
    \item The attacker probes all of the array to see which line was cached; the index of that line is the value of the secret data.
\end{enumerate}
The original Meltdown vulnerability targeted cache forwarding and allowed data extraction in this manner from \textit{any} memory address that was present in the cache.
\ac{MDS} attacks target smaller and more specific buffers in the on-core microarchitecture, and so make up a related but distinct class of attacks that are mitigated in a significantly different way.
While Meltdown targets the main memory that is updated relatively infrequently and shared across all cores, threads, and processes, \ac{MDS} attacks tend to target buffers that are local to cores (though sometimes shared across threads) and updated more frequently during execution.

\begin{figure*}[t]
    \centering
    \includegraphics[width=\linewidth]{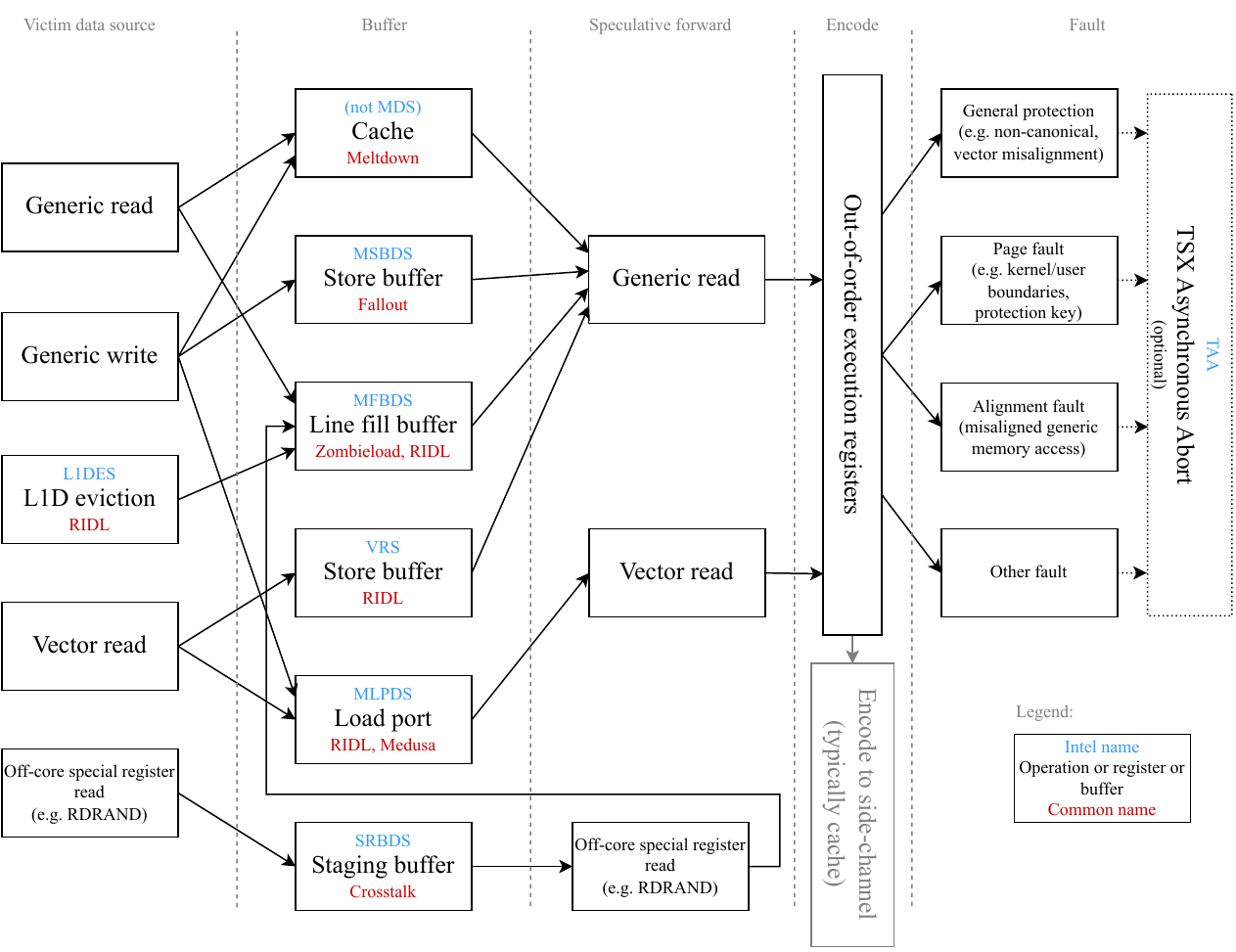}
    \caption{Major \acs{MDS} attack pathways and variant names on Intel CPUs. The blue names at the top are names of vulnerabilities given by Intel; the red names at the bottom are names given by researchers or the names of the papers in which the vulnerabilities were reported. Not all fault types work with all vulnerabilities on all systems---successful forwarding and encoding during speculative execution is dependent on the exact microarchitecture including any microarchitectural countermeasures that are in place, so cataloging every known combination would be beyond the scope of this paper.}
    \label{fig:mds-genealogy}
\end{figure*}

\subsubsection{Basic \acs{MDS} Variants}
\autoref{fig:mds-genealogy} charts the major known \ac{MDS} attack pathways on Intel CPUs and the names given to different variants by Intel and by the researchers who reported them.
Most broadly, Intel categorizes \ac{MDS} vulnerabilities in their CPUs by the specific buffer from which data is speculatively forwarded, since these buffers tend to be used for a number of different operations.
\ac{RIDL} \ac{MDS} vulnerabilities can be categorized as Microarchitectural Load Port Data Sampling (MLPDS), for variants that leak from the CPU's load port, and Microarchitectural Fill Buffer Data Sampling (MFBDS), for variants that leak from the CPU's \ac{LFB}.
Along the same lines, Intel calls the Fallout vulnerability Microarchitectural Store Buffer Data Sampling (MSBDS), as it involves a leakage from the store buffer.
Vector Register Sampling (VRS) is a variant of MSBDS that targets data that is handled by vector operations as it passes through the store buffer.
VERW bypass exploits a bug in the microcode fixes for MFBDS that loads stale and potentially secret data into the \ac{LFB}.
The basic mechanism of leakage is the same, and VERW bypass can be considered a special case of MFBDS.
L1 Data Eviction Sampling (L1DES) is another special case of MFBDS, where data that is evicted from the L1 data cache passes through the \ac{LFB} and becomes vulnerable to an \ac{MDS} attack. 
Notably, L1DES is a case where the attacker can actually trigger the secret data's presence in the CPU (by evicting it), whereas other \ac{MDS} attacks rely directly on the victim process accessing the secret data to bring it into the right CPU buffer. 

\subsubsection{Medusa}\label{sec:medusa-background}
Medusa~\cite{moghimi2020medusa} is a category of MDS attacks classified by Intel as MLPDS variants~\cite{intel2019mds}.
The Medusa vulnerabilities exploit the imperfect pattern-matching algorithms used to \textit{speculatively} combine stores in the write-combine (WC) buffer of Intel processors. 
Intel considers the WC buffer to be part of the load port, so Intel categorizes this vulnerability as a case of MLPDS.
There are three known Medusa variants which each exploit a different feature of the write-combine buffer to cause a speculative leakage:
\begin{description}
    \item[Cache Indexing:] a faulting load is speculatively combined with an earlier load with a matching cache line offset.
    \item[Unaligned Store-to-Load Forwarding:] a valid store followed by a dependent load that triggers an misaligned memory fault causes random data from the WC to be forwarded.
    \item[Shadow \repmov{}:] a faulting \repmov{} instruction followed by a dependent load leaks the data of a different \repmov{}.
\end{description}

\subsubsection{\acl{TAA}}
The hardware vulnerability \acf{TAA}~\cite{intel2019taa} provides a different speculation mechanism for carrying out an \ac{MDS} attack. 
While standard \ac{MDS} attacks access restricted data with a standard speculated execution, \ac{TAA} uses an atomic memory transaction as implemented by TSX. 
When an atomic memory transaction encounters an asynchronous abort, for example because another process reads a cache line marked for use by the transaction or because the transaction encounters a fault, all operations in the transaction are rolled back to the architectural state before the transaction started.
However, during this rollback, instructions inside the transaction that have already started execution can continue speculative execution, as in steps (2) and (3) of other \ac{MDS} attacks. 
\ac{TAA} impacts all Intel processors that support TSX, and the case of certain newer processors that are not affected by other \ac{MDS} attacks, \ac{MDS} mitigations or \ac{TAA}-specific mitigations (such as disabling TSX) must be implemented in software to protect against \ac{TAA}~\cite{intel2019taa}.

\subsubsection{Mitigations}
\label{sec:mds-mitigations}
Though Meltdown and \ac{MDS}-class vulnerabilities exploit low level microarchitectural operations, they can be mitigated with microcode firmware patches on most vulnerable CPUs.

\paragraph{Page table isolation} Historically, kernel page tables have been included in user-level process page tables so that a user-level process can make a system call to the kernel with minimal overhead. Page table isolation (first proposed by Gruss et al. as KAISER~\cite{gruss2019kaslr}) maps only the bare minimum necessary kernel memory into the user page table and introduces a second page table only accessible by the kernel. With the user process unable to access the kernel page table, accesses to all but a small and specifically chosen fraction of kernel memory are stopped before they reach the lower level caches where a Meltdown attack begins.

\paragraph{Buffer overwrite} \ac{MDS} attacks that target on-core CPU buffers require a lower-level and more targeted defense. 
Intel introduced a microcode update that overwrites vulnerable buffers when the first-level data (L1d) cache (a common target of cache timing side-channel attacks) is flushed or the \texttt{VERW} instruction is run~\cite{intel2019mds}. The kernel can then protect against \ac{MDS} attacks by triggering a buffer overwrite when switching to an untrusted process. 

The buffer overwrite mitigation targets \ac{MDS} attacks at their source, but is imperfect to say the least. 
Processes remain vulnerable to attacks from concurrently running threads on the same core when \ac{SMT} is enabled (since both threads share vulnerable buffers without the active process actually changing on either thread),
Furthermore, shortly after the original buffer overwrite microcode update, the \ac{RIDL} team found that on some Skylake CPUs, buffers were overwritten with stale and potentially sensitive data~\cite{vanschaik2019ridl}, and remained vulnerable even with mitigations enabled and \ac{SMT} disabled.
Still other processors are vulnerable to \ac{TAA} but not non-\ac{TAA} \ac{MDS} attacks, and did not receive a buffer overwrite microcode update and as such require that TSX be disabled completely to prevent \ac{MDS} attacks~\cite{intel2019taa,linux2020taa}.

\subsection{Spectre}
In 2018, Jan Horn and Paul Kocher~\cite{kocher2019spectre} independently reported the first Spectre variants. Since then, many different Spectre variants~\cite{kocher2019spectre,horn2018sectrev4,maisuradze2018ret2spec,koruyeh2018spectreReturns} and sub-variants~\cite{kiriansky2018speculativebufferoverflows,canela2019systematicevaluation,chen2019sgxpectre,barberis2022branchhistoryinjection,wikner2022retbleed} have been discovered. Spectre attacks make the CPU speculatively access memory that is architecturally inaccessible and leak the data into the architectural state. Therefore, all Spectre variants consist of three components~\cite{johannesmeyer2022kasper}:

The first component is the Spectre gadget that is speculatively executed. Spectre variants are usually separated by the source of the misprediction they exploit. The outcome of a conditional direct branch, e.g., is predicted by the Pattern History Table (PHT). Mispredictions of the PHT can lead to a speculative bounds check bypass for load and store instructions~\cite{kocher2019spectre,kiriansky2018speculativebufferoverflows,canela2019systematicevaluation}. The branch target of an indirect jump is predicted by the Branch Target Buffer (BTB). If an attacker can influence the result of a misprediction of the BTB, then speculative return-oriented programming attacks are possible~\cite{kocher2019spectre,canela2019systematicevaluation,chen2019sgxpectre,barberis2022branchhistoryinjection}. The same is true for predictions served by the Return Stack Buffer (RSB) that predicts return addresses during the execution of return instructions~\cite{maisuradze2018ret2spec,koruyeh2018spectreReturns,canela2019systematicevaluation}. Recent results showed that some modern CPUs use the BTB for their return address predictions if the RSB underflows~\cite{wikner2022retbleed}. Another source of Spectre attacks is the prediction of store-to-load dependencies. If a load is mispredicted to \emph{not} depend of a previous store, it speculatively executes on stale data which may lead to a speculative store bypass~\cite{horn2018sectrev4}. All of these gadgets are not exploitable by default but depend on the other two components discussed now.

The second component is how an attacker controls inputs to the aforementioned gadgets. Attackers may be able to define gadget input values directly through user input, file contents, network packets or other architectural mechanisms. On the other hand attackers may be able to inject data into the gadget transiently through load value injection~\cite{vanbulk2020lvi} or floating point value injection~\cite{ragab2021machineclear}. Attackers are able to successfully control gadget inputs if they can influence which data or instructions are accessed or executed during the speculation window.

The third component is the covert channel that is used to transfer the speculative microarchitectural state into an architectural state and therefore exfiltrate the speculatively accessed data into a persistent environment. Cache covert channels~\cite{yarom2014flushreload,osvik2006primeprobe,purnal2021primescope} are applicable if the victim code performs a transient memory access depending on speculatively accessed secret data~\cite{kocher2019spectre}. If a secret is accessed speculatively and loaded into an on-core buffer, an attacker can rely on an \ac{MDS}-based channel~\cite{canella2019fallout,schwarz2019zombieload,vanschaik2019ridl} to transiently transfer the exfiltrated data to the attacker thread where the data is transferred to the architectural state through, e.\,g., a cache covert channel. Last but not least, if the victim executes code depending on secret data, the attacker can learn the secret by observing port contention~\cite{bhattacharyya2019smotherspectre,aldaya2019portcontention,fustos2020spectrerewind,rokicki2022portcontentionwosmt,rokicki2022portableportcontention}.

\subsubsection{Mitigations}
Many countermeasures were developed to mitigate the various Spectre variants. A specific Spectre variant is effectively disabled if one of the three required components is removed. An attacker without control over inputs to Spectre gadgets is unlikely to successfully launch an attack. The same is true if a covert channel for transforming the speculative state into an architectural state is unavailable. But since this is usually hard to guarantee, Spectre countermeasures mainly focus on stopping mispredictions. Inserting \texttt{lfence} instructions before critical code sections disable speculative execution beyond this point and can therefore be used as a generic countermeasure. But because of its high performance overhead, more specific countermeasures were developed. Spectre-BTB countermeasures include Retpoline~\cite{google2018retpoline} and microcode updates like IBRS, STIBP, or IBPB~\cite{intel2018spectremitigation}. Spectre-RSB and Spectre-BTB-via-RSB can be mitigated by filling the RSB with values to overwrite malicious entries and prevent the RSB from underflowing or by installing IBRS microcode updates. Spectre-STL can be mitigated by the SSBD microcode update~\cite{intel2018spectremitigation}. Another drastic option to stop an attacker from tampering with shared branch prediction buffers is to disable \ac{SMT}. Disabling \ac{SMT} effectively partitions branch prediction hardware resources between concurrent tenants at the cost of a significant performance loss.

\subsection{\acs{AWS}'s isolation model}
\begin{figure}
    \centering
    \includegraphics[width=\linewidth]{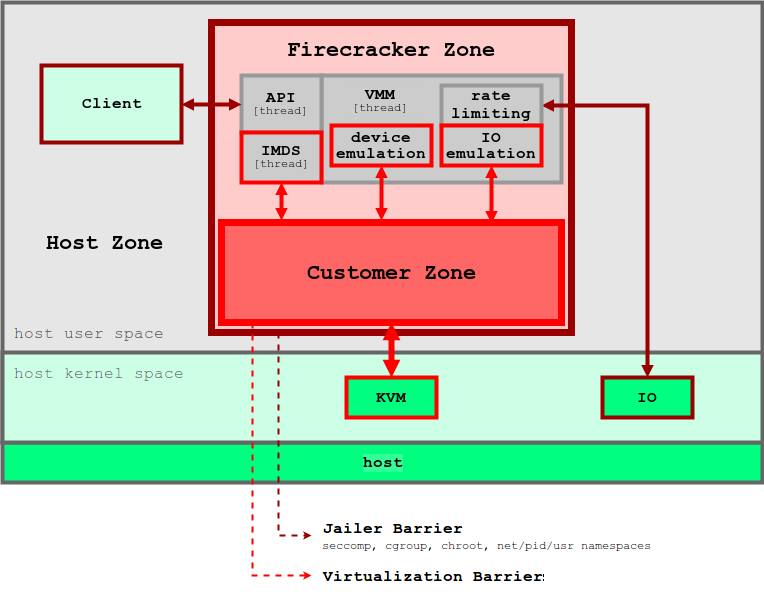}
    \caption{\acs{AWS} provides this threat containment diagram in a design document in the Firecracker GitHub repository~\cite{amazon2023firecrackerdesign}.
    In this model, the jailer provides container-like protections around Firecracker's \ac{VMM}, API, instance metadata service (IMDS), all of which run in the host user space, and the customer's workload, which runs inside the virtual machine.
    The \acs{VM} isolates the customer's workload in the guest, ensuring that it only directly interacts with predetermined elements of the host (in both user and kernel space).
    }
    \label{fig:firecracker-threat-containment}
\end{figure}

Firecracker is specifically built for serverless and container applications~\cite{agache2020firecracker} and is currently used by \ac{AWS}' Fargate \ac{CaaS} and Lambda \ac{FaaS}. 
In both of these service models, Firecracker is the primary isolation system that supports every individual Fargate task or Lambda event.
Both of these service models are also designed for running very high numbers of relatively small and short-lived tasks.
\ac{AWS} itemizes the design requirements for the isolation system that eventually became Firecracker as follows: 
\begin{displayquote}
\textbf{Isolation:} It must be safe for multiple functions to run on the
same hardware, protected against privilege escalation,
information disclosure, covert channels, and other risks.\\
\textbf{Overhead and Density:} It must be possible to run thousands of functions on a single machine, with minimal
waste.\\
\textbf{Performance:} Functions must perform similarly to running
natively. Performance must also be consistent, and isolated from the behavior of neighbors on the same hardware.\\
\textbf{Compatibility:} Lambda allows functions to contain arbitrary Linux binaries and libraries. These must be supported without code changes or recompilation.\\
\textbf{Fast Switching:} It must be possible to start new functions
and clean up old functions quickly.\\
\textbf{Soft Allocation:} It must be possible to over commit CPU,
memory and other resources, with each function consuming only the resources it needs, not the resources it is
entitled to.~\cite{agache2020firecracker}\\
\end{displayquote}
We are particularly interested in the \textbf{isolation} requirement and stress that microarchitectural attacks are declared \emph{in-scope} for the Firecracker threat model. 
The ``design'' page in \ac{AWS}'s public Firecracker Git repository elaborates on the isolation model and provides a useful diagram which we reproduce in \autoref{fig:firecracker-threat-containment}. 
This diagram pertains mostly to protection against privilege escalation.
The outermost layer of protection is the jailer, which uses container isolation techniques to limit the Firecracker's access to the host kernel while running the \ac{VMM} and other management components of Firecracker as threads of a single process in the host userspace.
Within the the Firecracker process, the user's workload is run on other threads.
The workload threads execute the guest operating system of the virtual machine and any programs running in the guest.
Running the user's code in the virtual machine guest restricts its direct interaction with the host to prearranged interactions with \ac{KVM} and certain portions of the Firecracker management threads.
So from the perspective of the host kernel, the \ac{VMM} and the \ac{VM} including the user's code are run in the same process.
This is the reason why \ac{AWS} states that each \ac{VM} resides in a single process.
But, since the \ac{VM} is isolated via hardware virtualization techniques, the user's code, the guest kernel, and the \ac{VMM} operate in separate address spaces.
Therefore, the guest's code cannot architecturally or transiently access \ac{VMM} or guest kernel memory addresses as they are not mapped in the guest's address space.
The remaining microarchitectural attack surface is limited to \ac{MDS} attacks that leak information from CPU internal buffers ignoring address space boundaries and Spectre attacks where an attacker manipulates the branch prediction of other processes to self-leak information.

Not shown in \autoref{fig:firecracker-threat-containment}, but equally important to \ac{AWS}'s threat model, is the \textbf{isolation} of functions from each other when hardware is shared, especially in light of the \textbf{soft allocation} requirement.
Besides the fact that compromising the host kernel could compromise the security of any guests, microarchitectural attacks that target the host hardware can also threaten user code directly.
Since a single Firecracker process contains all the necessary threads to run a virtual machine with a user's function, soft allocation can simply be performed by the host operating system~\cite{agache2020firecracker}.
This means that standard Linux process isolation systems are in place on top of virtual machine isolation.

\subsubsection{Firecracker security recommendations}
\label{sec:firecracker-security-recs}
The Firecracker documentation also recommends the following precautions for protecting against microarchitectural side-channels~\cite{amazon2023prodhostsetup}:
\begin{itemize}
    \item Disable \ac{SMT}
    \item Enable kernel page-table isolation
    \item Disable kernel kame-page merging
    \item Use a kernel compiled with Spectre-BTB mitigation (e.g., IBRS and IBPB on x86)
    \item Verify Spectre-PHT mitigation 
    \item Enable L1TF mitigation
    \item Enable Spectre-STL mitigation
    \item Use memory with Rowhammer mitigation
    \item Disable swap or use secure swap
\end{itemize}



%





\section{Threat models}
We propose two threat models applicable to Firecracker-based serverless cloud systems:

\begin{enumerate}
    \item 
The \emph{user-to-user} model (\autoref{fig:user-user}): a malicious user runs arbitrary code sandboxed within a Firecracker \ac{VM} and attempts to leak data, inject data, or otherwise gain information about or control over another user's sandboxed application.
In this model, we consider 
\begin{enumerate}
    \item the time-sliced sharing of hardware, where the instances of the two users execute in turns on the CPU core, and 
    \item physical co-location, where the two users' code runs concurrently on hardware that is shared in one way or another (for example, two cores on the same CPU or two threads in the same core if \ac{SMT} is enabled). 
\end{enumerate}

\item
The \emph{user-to-host} model (\autoref{fig:user-host}): a malicious user targets some component of the host system: the Firecracker \ac{VMM}, \ac{KVM}, or another part of the host system kernel. For this scenario, we only consider time-sliced sharing of hardware resources. This is because the host only executes code if the guest user's \ac{VM} exits, e.\,g. due to a page fault that has to be handled by the host kernel or \ac{VMM}.

 \begin{figure}
     \centering
     \begin{tikzpicture}[
node distance=0cm,
font={\small},
alignc/.style={align=center},
alignl/.style={align=left},
alignr/.style={align=right},
high/.style={minimum height=.65cm},
small dim/.style={high, minimum width=.225\linewidth, text width=.195\linewidth},
wide dim/.style={high, minimum width=.45\linewidth, text width=.42\linewidth},
medium wide dim/.style={high, minimum width=.675\linewidth, text width=.645\linewidth},
very wide dim/.style={high, minimum width=.9\linewidth, text width=.87\linewidth},
green/.style={draw=nicegreenborder, fill=nicegreenfill, very thick},
blue/.style={draw=niceblueborder, fill=nicebluefill, very thick},
red/.style={draw=niceredborder, fill=niceredfill, text=white, very thick},
orange/.style={draw=niceorangeborder, fill=niceorangefill, very thick},
yellow/.style={draw=niceyellowborder, fill=niceyellowfill, very thick},
]

\node[very wide dim, alignc, blue] (cpu) {CPU};
\node[very wide dim, alignc, orange, above=of cpu.north west, anchor=south west] (kernel) {Host Kernel};

\node[wide dim, alignc, orange, above=of kernel.north west, anchor=south west] (kvm0) {KVM};
\node[wide dim, alignc, yellow, above=of kvm0.north west, anchor=south west] (fc0) {Firecracker};
\node[wide dim, alignc, yellow, above=of fc0.north west, anchor=south west] (kernel0) {Guest Kernel};
\node[wide dim, alignc, red, above=of kernel0.north west, anchor=south west] (rt0) {Runtime};
\node[wide dim, alignc, red, above=of rt0.north west, anchor=south west] (app0) {App};
\node[devil,minimum size=.75cm,anchor=west] at (rt0.north west) {};

\node[wide dim, alignc, orange, above=of kernel.north east, anchor=south east] (kvm1) {KVM};
\node[wide dim, alignc, yellow, above=of kvm1.north east, anchor=south east] (fc1) {Firecracker};
\node[wide dim, alignc, yellow, above=of fc1.north east, anchor=south east] (kernel1) {Guest Kernel};
\node[wide dim, alignc, yellow, above=of kernel1.north west, anchor=south west] (rt1) {Runtime};
\node[wide dim, alignc, green, above=of rt1.north east, anchor=south east] (app1) {App};
\node[alice,good,minimum size=.75cm,anchor=east] at (app1.north east) {};

\draw[->,draw=niceredborder] plot [smooth] coordinates {([yshift=-5]rt0) (cpu) ([yshift=-5]app1)};
\end{tikzpicture}
     \caption{In the \emph{user-to-user} threat model, we assume that a malicious cloud service tenant attempts to ex-filtrate information from another tenant. We assume the attacker to have control over the app and runtime of its \acs{VM} while the guest kernel is provided by the \acs{CSP}.}
     \label{fig:user-user}
 \end{figure}
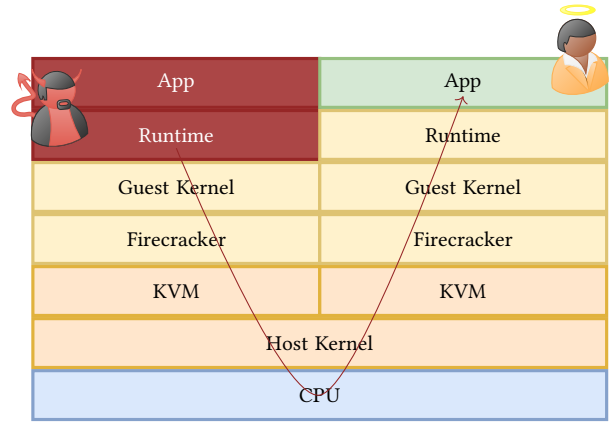
 \begin{figure}
     \centering
     \begin{tikzpicture}[
node distance=0cm,
font={\small},
alignc/.style={align=center},
alignl/.style={align=left},
alignr/.style={align=right},
high/.style={minimum height=.65cm},
small dim/.style={high, minimum width=.225\linewidth, text width=.195\linewidth},
wide dim/.style={high, minimum width=.45\linewidth, text width=.42\linewidth},
medium wide dim/.style={high, minimum width=.675\linewidth, text width=.645\linewidth},
very wide dim/.style={high, minimum width=.9\linewidth, text width=.87\linewidth},
green/.style={draw=nicegreenborder, fill=nicegreenfill, very thick},
blue/.style={draw=niceblueborder, fill=nicebluefill, very thick},
red/.style={draw=niceredborder, fill=niceredfill, text=white, very thick},
orange/.style={draw=niceorangeborder, fill=niceorangefill, very thick},
yellow/.style={draw=niceyellowborder, fill=niceyellowfill, very thick},
]

\node[very wide dim, alignc, blue] (cpu) {CPU};
\node[very wide dim, alignc, green, above=of cpu.north west, anchor=south west] (kernel) {Host Kernel};

\node[wide dim, alignc, green, above=of kernel.north west, anchor=south west] (kvm0) {KVM};
\node[wide dim, alignc, green, above=of kvm0.north west, anchor=south west] (fc0) {Firecracker};
\node[wide dim, alignc, yellow, above=of fc0.north west, anchor=south west] (kernel0) {Guest Kernel};
\node[wide dim, alignc, red, above=of kernel0.north west, anchor=south west] (rt0) {Runtime};
\node[wide dim, alignc, red, above=of rt0.north west, anchor=south west] (app0) {App};
\node[devil,minimum size=.75cm,anchor=west] at (rt0.north west) {};

\node[wide dim, alignc, orange, above=of kernel.north east, anchor=south east] (kvm1) {KVM};
\node[wide dim, alignc, yellow, above=of kvm1.north east, anchor=south east] (fc1) {Firecracker};
\node[wide dim, alignc, yellow, above=of fc1.north east, anchor=south east] (kernel1) {Guest Kernel};
\node[wide dim, alignc, yellow, above=of kernel1.north west, anchor=south west] (rt1) {Runtime};
\node[wide dim, alignc, yellow, above=of rt1.north east, anchor=south east] (app1) {App};
\node[alice,good,minimum size=.75cm,anchor=west] at (kvm0.west) {};

\draw[->,draw=niceredborder] ([yshift=-55]rt0) edge [bend right] ([yshift=5]kernel);
\end{tikzpicture}
     \caption{In the \emph{user-to-host} threat model, the malicious tenant aims for information ex-filtration from the host system, e.\,g. the \acl{VMM} or the host kernel. The attacker has control over the runtime and app in its \acl{VM} while the guest kernel is provided by the \ac{CSP}.}
     \label{fig:user-host}
 \end{figure}
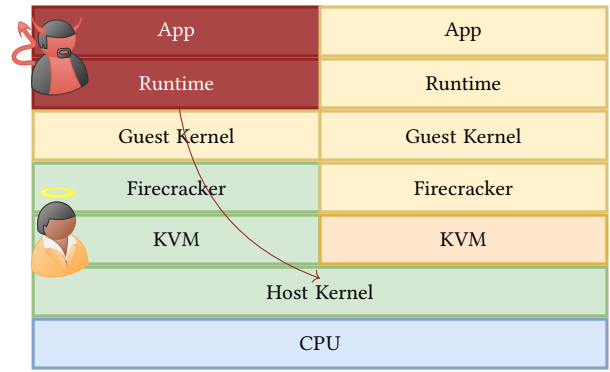
\end{enumerate}

For both models, we assume that a malicious user is able to control the runtime environment of its application.
In our models, malicious users do \emph{not} posses guest kernel privileges. 
Therefore, both models grant the attacker slightly less privileges than the model assumed by~\cite{agache2020firecracker} where the guest kernel is chosen and configured by the \ac{VMM} but assumed to be compromised at runtime.
Rather, the attacker's capabilities in our models match the capabilities granted to users in deployments of Firecracker in \acs{AWS} Lambda and Fargate.

\section{Analysis of Firecracker's Containment Systems}
\autoref{fig:firecracker-threat-containment} shows the containment offered by Firecracker, as presented by AWS. In this section, we analyze each depicted component and their defenses against and vulnerabilities to microarchitectural attacks.

\paragraph{KVM}
Linux \acf{KVM} is the hypervisor implemented in modern Linux kernels and therefore part of the Linux code base.
It virtualizes the supervisor and user modes of the underlying hardware, manages context switches between \acp{VM}, and handles most VM-exit reasons unless they are related to I/O operations.
Besides these architectural isolation mechanisms, \ac{KVM} also implements mitigations against Spectre attacks on a VM-exit to protect the host \ac{OS} or hypervisor from malicious guests.
Firecracker heavily relies on \ac{KVM} as its hypervisor.
However, since \ac{KVM} is part of the Linux source code and developed by the Linux community, we define \ac{KVM} to not be a part of Firecracker.
Therefore, countermeasures against microarchitectural attacks that are implemented in \ac{KVM} cannot be attributed to Firecracker's containment system.

\paragraph{Metadata, device, and I/O services}
The metadata, device, and I/O services are the parts of the Firecracker \ac{VMM} and API that interact directly with a \ac{VM}, collecting and managing metrics and providing connectivity.
AWS touts the simplicity of these interfaces (for a reduced attack surface) and that they are written from scratch for Firecracker in Rust, a programming language known for its security features~\cite{balasubramanian2017system}.
However, Rust most notably provides in-process protection against invalid and out-of-bounds memory accesses, but microarchitectural attacks like cache attacks, Spectre, and \ac{MDS} can leak information between processes rather than directly hijacking a victim's process. 

Another notable difference between Firecracker and many other \acp{VMM} is that all of these services are run within the same host process as the \ac{VM} itself, albeit in another thread. 
While the virtualization of memory addresses within the \ac{VM} provides some obfuscation between the guest's code and the I/O services, some Spectre attacks work specifically within a single process.
Intra-process attacks may pose less of a threat to real world systems, however, since two guests running on the same hardware each have their own copy of these essential services.

\paragraph{Jailer barrier}
In the event that the API or \ac{VMM} are compromised, the jailer provides one last barrier of defense around a Firecracker instance.
It protects the host system's files and resources with namespaces and control groups (cgroups), respectively~\cite{amazon2023jailer}.
Microarchitectural attacks do not directly threaten files, which are by definition outside of the microarchitectural state.
Cgroups allow a system administrator to assign processes to groups and then allocate, constrain, and monitor system resource usage on a per-group basis~\cite{redhat2020resource}.
It is plausible that limitations applied with cgroups could impede an attacker's ability to carry out certain microarchitectural attacks.
For example, memory limitations might make it difficult to carry out eviction-based cache attacks, or CPU time limitations could prevent an attacker from making effective use of a CPU denial-of-service tool like HyperDegrade~\cite{aldaya2022hyperdegrade} which can slow down a victim process, simplifying the timing of a microarchitectural side-channel exfiltration or injection.
In practice, Firecracker is not distributed with any particular cgroup rules~\cite{amazon2023jailer}; in fact, it is specifically designed for the efficient operation of many Firecracker \acp{VM} under the default Linux resource allocation~\cite{amazon2023firecrackerdesign}.

None of the isolation and containment systems in Firecracker seem to directly protect against user-to-user or user-to-host attacks. 
Therefore, we proceeded to test various microarchitectural attack proof of concepts inside and outside of Firecracker VMs.

\section{Analysis of microarchitectural attacks and defenses in Firecracker microVMs}
In this section we present our analysis of a number of microarchitectural side-channel and speculative attack \acp{PoC} on Firecracker microVMs.
We test these \acp{PoC} on bare metal and in Firecracker instances, and test relevant microcode defenses in the various scenarios.
We run our tests on a server with an Intel Skylake 4114 CPU which has virtualization hardware extensions and \ac{SMT} enabled.
The CPU runs on microcode version 0x2006b06\footnote{Updating the microcode to a newer version would disable TSX on our system which would make tests with TSX-based \ac{MDS} variants impossible.}.
The host \ac{OS} is Ubuntu 20.04 with a Linux 5.10 kernel.
We used Firecracker v1.0.0 and v1.4.0, the latest version as of July 2023, to run an Ubuntu 18.04 guest with Linux kernel 5.4 which is provided by Amazon when following the quick-start guide\footnote{\url{https://github.com/firecracker-microvm/firecracker/blob/dbd9a84b11a63b5e5bf201e244fe83f0bc76792a/docs/getting-started.md}}.

In summary, the recommended production host setup provided with \ac{AWS} Firecracker is insufficient when it comes to protecting tenants from malicious neighbors. Firecracker therefore fails in providing its claimed isolation guarantees. This is because

\begin{enumerate}
    \item we identify a Medusa variant that only becomes exploitable when it is run across microVMs. In addition, the recommended countermeasures do not contain the necessary steps to mitigate the side-channel, or most other Medusa variants.
    \item we show that tenants are not properly protected from information leaks induced through Spectre-PHT or Spectre-BTB when applying the recommended countermeasures. The Spectre-PHT variants remain a problem even when disabling \ac{SMT}.
    \item we observed no differences in \ac{PoC} performance between Firecracker v1.0.0 and v1.4.0.
\end{enumerate}
 We conclude that the virtualization layer provided by Firecracker has little effect on microarchitectural attacks, and Firecracker's system security recommendations are insufficient.

\subsection{Medusa}

\begin{table}[t]
    \centering
    \caption{Presence of Medusa side-channels with all microarchitectural defense kernel options disabled. Note that the combination of cache indexing leak and block write secret (highlighted in yellow) works in Firecracker \acsp{VM} but not on bare metal.}
    \label{medusa-all-disabled-2}
    \newcommand{\mycc}{\cellcolor{niceyellowfill}}
\begin{tabular}{c c c c c }
    \toprule
    \textbf{Leak}                                          &                          & \textbf{Secret}   & \textbf{Bare Metal} & \textbf{Firecracker} \\
    \midrule
    \multirow{2}{*}{Cache Indexing}                        & \ldelim\{{2}[.25em]{*}[] & \mycc Block Write & \mycc\nochannel     & \mycc\channel        \\
                                                           &                          & \repmov           & \channel            & \channel           \\
    \addlinespace[.5em]
    \multirow{2}{*}{\shortstack{Unaligned\\Store-to-Load}} & \ldelim\{{2}[.25em]{*}[] & Block Write       & \channel            & \channel             \\
                                                           &                          & \repmov           & \channel            & \channel             \\
    \addlinespace[.5em]
    \multirow{2}{*}{\shortstack{Shadow \\ \repmov}}        & \ldelim\{{2}[.25em]{*}[] & Block Write       & \nochannel          & \nochannel           \\
                                                           &                          & \repmov           & \channel            & \channel             \\
    \midrule
    \multicolumn{5}{p{.9\linewidth}}{\channelkey}
\end{tabular}
\end{table}

\begin{table}[t]
    \centering
    \caption{Mitigations necessary to protect bare metal vs. Firecracker victims from Medusa attacks. Note that \acs{AWS}'s recommended mitigation, \texttt{nosmt}, does not prevent the highlighted cache indexing/block write variant that is enabled by Firecracker (cf. \autoref{medusa-all-disabled-2}), or any other variant except shadow \repmov{}.}
    \label{medusa-mininimum-protections}
\newcommand{\mycc}{\cellcolor{niceyellowfill}}
\newcommand{\mitspace}{\hspace{4.5pt}}
\centering
\begin{tabular}{@{}c@{}c@{\hspace{5pt}}c c@{\mitspace} c c @{\mitspace} c @{\mitspace} c}
    \toprule
     & &  & \multicolumn{2}{c}{\textbf{Bare metal}} & \multicolumn{3}{c}{\textbf{Firecracker}} \\
    \cmidrule(rl){4-5} \cmidrule(rl){6-8}
    \textbf{Leak} & & \textbf{Secret} & \texttt{mds} & \texttt{nosmt} & \texttt{mds}(VM) & \texttt{mds}(H) & \texttt{nosmt} \\
    \midrule
    \multirow{2}{*}{\shortstack{Cache\\Indexing}} & \ldelim\{{2}[.25em]{*}[] & \mycc Block Write & \mycc\novuln & \mycc\novuln & \mycc\noprotect &\mycc\protects & \mycc\noprotect \\
     & & \repmov & \protects & \protects & \protects & \protects & \protects \\ [.5ex]
    \addlinespace[.5em]
    \multirow{2}{*}{\shortstack{Unaligned\\Store}} & \ldelim\{{2}[.25em]{*}[] & Block Write & \protects & \noprotect & \noprotect & \protects & \noprotect \\
     & & \repmov & \protectswith & \protectswith & \noprotect & \protectswith & \protectswith \\
    \addlinespace[.5em]
    \multirow{2}{*}{\shortstack{Shadow \\\repmov}} & \ldelim\{{2}[.25em]{*}[] & Block Write & \novuln & \novuln & \novuln & \novuln & \novuln \\
     & & \repmov & \protects & \protects & \protectswith & \protects & \protectswith \\
     \midrule
    \multicolumn{7}{p{\columnwidth-1cm}}{ 
    \protectskey\newline 
    }
\end{tabular}
\end{table}

\begin{table*}[t]
    \centering
    \caption{Mitigations necessary to protect bare metal vs. Firecracker victims from \acs{RIDL} and other \ac{MDS} attacks. The recommended \texttt{nosmt} mitigation protects against most but not all of these variants. All proof of concepts were tested on Firecracker v1.0.0 and v1.4.0 with identical results.}
    \label{ridl-mininimum-protections}
\begin{tabular}{ c ccc cc ccc c }

\toprule
\multirow{2}{*}{\textbf{Exploit}}
    & \multicolumn{3}{c}{\textbf{Details}}
        & \multicolumn{2}{c}{\textbf{Bare Metal}}
            & \multicolumn{3}{c}{\textbf{Firecracker}}
                & \\
    \cmidrule(rl){2-4}
        \cmidrule(rl){5-6}
            \cmidrule(rl){7-9}
    & Common Name               
    & Target Buffer             & Fault Type
        &\texttt{nosmt}             & \texttt{mds}
            & \texttt{nosmt} (H)        & \texttt{mds} (H)  & \texttt{mds} (VM)
                & \multirow[t]{ 2}{*}{\shortstack{TSX\\required?}} \\
\midrule                                                  
\texttt{alignment\_write}
    & \multirow{4}{*}{RIDL}     
    & Fill Buffer               & Alignment
        & \protectswith             & \protectswith
            & \protectswith             & \protectswith     & \noprotect
                & no \\

\texttt{pgtable\_leak\_notsx}
    &                           
    & Fill Buffer               & Page
        & \noprotect$^b$            & \protects
            & \noprotect$^b$            & \protects         & \protects
                & no \\
\texttt{ridl\_basic}
    &                           
    & Fill Buffer               & Page
        & \protects                 & \noprotect
            & \novuln$^c$               & \novuln$^c$       & \novuln$^c$
                & no \\
\texttt{ridl\_invalidpage}
    &                           
    & Fill Buffer               & Page
        & \protects                 & \noprotect
            & \novuln$^c$               & \novuln$^c$       & \novuln$^c$
                & no \\
\midrule
\texttt{pgtable\_leak}
    & \multirow{3}{*}{RIDL/TAA} 
    & Fill Buffer               & TSX abort
        & \protects                 & \noprotect
            & \protects                 & \noprotect        & \noprotect
                & yes \\
\texttt{taa\_read}
    &                           
    & Fill Buffer               & TSX abort
        & \protects                 & \noprotect
            & \protects                 & \noprotect        & \noprotect
                & yes \\
\texttt{taa\_basic}
    &                           
    & Fill Buffer               & TSX abort
        & \protects                 & \noprotect
            & \novuln$^c$               & \novuln$^c$       & \novuln$^c$
                & yes \\
\midrule
\texttt{verw\_bypass\_l1des}
    & RIDL/TAA                  
    & Fill Buffer               & TSX abort
        & \protects                 & \noprotect
            & \protects                 & \noprotect        & \noprotect
                & yes \\
\texttt{loadport}
    & RIDL                      
    & Load Port                 & Page
        & \protects                 & \noprotect
            & \protects                 & \noprotect        & \noprotect
                & no \\
\texttt{vrs}
    & RIDL/VRS                  
    & Store buffer              & Page
        & \protects                 & \noprotect
            & \protects                 & \noprotect        & \noprotect
                & no \\
\texttt{cpuid\_leak}
    & Crosstalk                 
    & Fill Buffer               & Page
        & \protects                 & \noprotect
            & \novuln$^a$               & \novuln$^a$       & \novuln$^a$
                & no \\
\bottomrule
\multicolumn{10}{p{\textwidth - 10pt}}{\protectskey} \\
\multicolumn{10}{l}{$^a$ \texttt{CPUID} instruction is emulated by virtual machine and has no microarchitectural effect.} \\
\multicolumn{10}{l}{$^b$ This attack leaks information about pages used in its own thread.} \\
\multicolumn{10}{p{\textwidth - 10pt}}{$^c$ PoCs had to be modified slightly to run in two processes before they could be tested in the virtual machine. These PoCs did not work on bare metal or in virtual machines when split into two processes.}
\end{tabular} 

\end{table*}


We evaluated Moghimi's \acp{PoC}~\cite{moghimi2020medusa-code} for the Medusa~\cite{moghimi2020medusa} side-channels (classified by Intel as MLPDS variants of \ac{MDS}~\cite{intel2019mds}) on the bare metal of our test system and in Firecracker VMs.
There is one leaking \ac{PoC} for each of the three known variants described in \autoref{sec:medusa-background}.
We used two victim programs from the \ac{PoC} library: 
\begin{itemize}
    \item The ``Block Write'' program writes a large amount of consecutive data in a loop (so that the processor will identify repeated stores and combine them). 
    \item The ``\repmov{}'' program performs a similar operation, but with the \repmov{} instruction instead of many instructions moving smaller blocks of data in a loop.
\end{itemize}

\subsubsection{Results}
\autoref{medusa-all-disabled-2} shows the cases in which data is successfully leaked with all microarcitectural protections in the kernel disabled.
The left two columns show the possible combinations of the three Medusa \acp{PoC} and the two included victim programs. 
The right columns indicate which configurations work on bare metal and with the secret and leaking program running in parallel Firecracker instances. 
Most notably, with the Cache Indexing variant, the Block Write secret only works with Firecracker.
This is likely because of the memory address virtualization that the virtual machine provides: the guest only sees virtual memory regions mapped by \ac{KVM}, and \ac{KVM} traps memory access instructions and handles the transactions on behalf of the guest.

We tested the effectiveness of \texttt{mds} and \texttt{nosmt} defenses against each combination of attacker and victim \ac{PoC} on bare metal and in Firecracker \acp{VM}.
\autoref{medusa-mininimum-protections} lists the protections necessary to prevent Medusa attacks in bare metal and Firecracker scenarios.
Across the four vulnerabilities in Firecracker, only one is mitigated by \texttt{nosmt} alone, and \ac{AWS} does \emph{not explicitly recommend} enabling the \texttt{mds} protection, though most Linux distributions ship with it enabled by default.
That is to say, a multi-tenant cloud platform could be using Firecracker in a way that is compliant with \ac{AWS}'s recommended security measures and still be vulnerable to the majority of Medusa variants, including one where the Firecracker \ac{VMM} itself leaks the user's data that would not otherwise be leaked.

\subsection{\acs{RIDL} and More}
In this section, we present an evaluation of the \ac{RIDL} \ac{PoC} programs~\cite{vanschaik2020ridl-code} provided alongside van Schaik et al.'s 2019 paper~\cite{vanschaik2019ridl}.
\ac{RIDL} is a class of \ac{MDS} attacks that exploits speculative loads from buffers inside the CPU (not from cache or memory). 
The \ac{RIDL} \ac{PoC} repository also includes examples of attacks released in later addenda to the paper as well as one variant of the Fallout \ac{MDS} attack.

\subsubsection{Results}
\autoref{ridl-mininimum-protections} shows some basic information about the \ac{RIDL} \acp{PoC} that we tested and the efficacy of relevant countermeasures at preventing the attacks.
We compared attacks on bare metal and in Firecracker to evaluate Amazon's claims of the heightened hardware security of the Firecracker microVM system. 
For tests on the Firecracker system, we distinguish between countermeasure flags enabled on the host system (H) and the Firecracker guest kernel (\ac{VM}).
Besides the \texttt{nosmt} and \texttt{mds} kernel flags, we tested other relevant flags (cf. \autoref{sec:mds-mitigations},~\cite{linux2019mds}), including \texttt{kaslr}, \texttt{pti}, and \texttt{l1tf}, but did not find that they had an affect on any of these programs.
We excluded the \texttt{tsx\_async\_abort} mitigation since the CPU we tested on includes \texttt{mds} mitigation which makes the \texttt{tsx\_async\_abort} kernel flag redundant~\cite{linux2020taa}.

In general, we found that the \texttt{mds} protection does not adequately protect against the majority of \ac{RIDL} attacks.
However, disabling \ac{SMT} does mitigate the majority of these exploits.
This is consistent with Intel's~\cite{intel2019mds} and the Linux developers'~\cite{linux2019mds} statements that \ac{SMT} must be disabled to prevent \ac{MDS} attacks across hyperthreads.
The two outliers among these \acp{PoC} are \texttt{alignment\_write}, which requires both \texttt{nosmt} and \texttt{mds} on the host,
and \texttt{pgtable\_leak\_notsx}, which is mitigated only by \texttt{mds} countermeasures.
The leak relying on \texttt{alignment\_write} uses an alignment fault rather than a page table fault leak to trigger speculation~\cite{vanschaik2019ridl}.
The \ac{RIDL} paper~\cite{vanschaik2019ridl} and Intel's documentation of the related VRS exploit~\cite{intel2020vrs} are unclear about what exactly differentiates this attack from the page-fault-based MFBDS attacks found in other \acp{PoC}, but our experimental findings indicate that the microarchitectural mechanism of the leakage is distinct.
There is a simple and reasonable explanation for the behavior of \texttt{pgtable\_leak\_notsx}, which is unique among these PoCs for one key reason: it is the only exploit that crosses security boundaries (leaking page table values from the kernel) within a single thread rather than leaking from another thread.
It is self-evident that disabling multi-threading would have little effect on this single-threaded exploit. 
However, the \texttt{mds} countermeasure flushes microarchitectural buffers before switching from kernel-privilege execution to user-privilege execution within the same thread, wiping the page table data accessed by kernel code from the \ac{LFB} before the attacking user code can leak it.

In contrast to Medusa, most of these PoCs are mitigated by \ac{AWS}'s recommendation of disabling \texttt{smt}. 
However, as with Medusa, the Firecracker \ac{VMM} itself provides no microarchitectural protection against these attacks.

\subsection{Spectre}
Next we focus on Spectre vulnerabilities.
While there have been many countermeasures developed since Spectre attacks were first discovered, many of them either come with a (significant) performance penalty or only partially mitigate the attack.
Therefore, system operators often have to decide for a performance vs. security trade-off.
In this section we evaluate the Spectre attack surface available to Firecracker tenants in both threat models described earlier.
To evaluate the wide range of Spectre attacks, we rely on the \acp{PoC} provided in~\cite{canela2019transientfail-code}.
For Spectre-PHT, Spectre-BTB, and Spectre-RSB, the repository contains four PoCs each.
They differ in the way the attacker mistrains the \ac{BPU}.
The four possibilities are (1) \emph{same-process}--the attacker has control over the victim process or its inputs to mistrain the \ac{BPU}--(2) \emph{cross-process}--the attacker runs its own code in a separate process to influence the branch predictions of the victim process--(a) \emph{in-place}--the attacker mistrains the the \ac{BPU} with branch instruction that reside at the same virtual address as the target branch that the attacker wants to be misspredicted in the victim process--(b) \emph{out-of-place}--the attacker mistrains the \ac{BPU} with branch instructions that reside at addresses that are congruent to the target branches in the victim process.
\begin{enumerate}
    \item \emph{same-process:} the attacker has control over the victim process or its inputs to mistrain the \ac{BPU},
    \item \emph{cross-process:} the attacker runs its own code in a separate process to influence the branch predictions of the victim process,
    \item \emph{in-place:} the attacker mistrains the the \ac{BPU} with branch instruction that reside at the same virtual address as the target branch that the attacker wants to be misspredicted in the victim process
    \item \emph{out-of-place:} the attacker mistrains the \ac{BPU} with branch instructions that reside at addresses that are congruent to the target branches in the victim process.
\end{enumerate}
The first two and latter two situations are orthogonal, so each PoC combines two of them. For Spectre-STL, only same-process variants are known, which is why the repository only provides two PoCs in this case.
For cross-VM experiments, disabled address space layout randomization for the host and guest kernels as well as for the host and guest user level to ease finding congruent addresses that are used for mistraining.

\subsubsection{Results}

\begin{table*}[t]
    \centering
    \caption{Spectre \acsp{PoC} run with \ac{AWS} Firecracker recommended countermeasures (cf. \autoref{lst:enabled-spectre-mitigations} and~\cite{amazon2023prodhostsetup}). These countermeasures---which are the default for the Linux kernels in use---are insufficient when it comes to protecting tenants from Spectre attacks. Experiments with Firecracker v1.0.0 and v1.4.0 yielded the same results.}
    \label{tab:spectre-pocs-protected}
    \newcommand{\diff}{}
\newcommand{\host}{\faServer}
\newcommand{\fire}{\faFire~}
\newcommand{\crossfire}{\faFire$\leftrightarrow$\faFire~}

\begin{tabular}{c c c cc cc}
\toprule
\multirow{2}{*}{Variant}
                     & \multirow{2}{*}{Platform}
                                                          & \multirow{2}{*}{Thread Configuration}
                                                                                     & \multicolumn{2}{c}{same-process} & \multicolumn{2}{c}{cross-process} \\
                                                                                       \cmidrule(rl){4-5}                 \cmidrule(rl){6-7}
                     &                                    &                          & in-place   & out-of-place        & in-place   & out-of-place         \\
\midrule
\multirow{7}{*}{PHT} & \multirow{1}{*}{\host}             & any                      & \noprotect & \noprotect          & \noprotect & \noprotect           \\
                     & \multirow{1}{*}{\fire v1.0.0}      & any                      & \noprotect & \noprotect          & \noprotect & \noprotect           \\
                     & \multirow{1}{*}{\fire v1.4.0}      & any                      & \noprotect & \noprotect          & \noprotect & \noprotect           \\
                       \cmidrule(rl){3-7}
                     & \multirow{2}{*}{\crossfire v1.0.0} & \pinned(1PT), 1vCPU each & \novuln    & \novuln             & \noprotect & \noprotect           \\
                     &                                    & \pinned(2PT), 1vCPU each & \novuln    & \novuln             & \noprotect & \protects            \\
                       \cmidrule(rl){3-7}
                     & \multirow{2}{*}{\crossfire v1.4.0} & \pinned(1PT), 1vCPU each & \novuln    & \novuln             & \noprotect & \noprotect           \\
                     &                                    & \pinned(2PT), 1vCPU each & \novuln    & \novuln             & \noprotect & \protects            \\
\midrule
\multirow{12}{*}{BTB} & \multirow{3}{*}{\host}            & \pinned(1PT)             & \noprotect & \protects           & \noprotect & \protects            \\
                     &                                    & \pinned(2PT)             & \noprotect & \protects           & \noprotect & \protects            \\
                       \cmidrule(rl){3-7}
                     & \multirow{3}{*}{\fire v1.0.0}      & \pinned(1PT), 1vCPU      & \noprotect & \protects           & \noprotect & \protects            \\
                     &                                    & \pinned(1PT), 2vCPU      & \noprotect & \protects           & \diff\protects  & \protects            \\
                     &                                    & \pinned(2PT), 2vCPU      & \noprotect & \protects           & \noprotect & \protects            \\
                       \cmidrule(rl){3-7}
                     & \multirow{3}{*}{\fire v1.4.0}      & \pinned(1PT), 1vCPU      & \noprotect & \protects           & \noprotect & \protects            \\
                     &                                    & \pinned(1PT), 2vCPU      & \noprotect & \protects           & \diff\protects  & \protects            \\
                     &                                    & \pinned(2PT), 2vCPU      & \noprotect & \protects           & \noprotect & \protects            \\
                       \cmidrule(rl){3-7}
                     & \multirow{2}{*}{\crossfire v1.0.0} & \pinned(1PT), 1vCPU each & \novuln    & \novuln             & \protects  & \protects            \\
                     &                                    & \pinned(2PT), 1vCPU each & \novuln    & \novuln             & \noprotect & \protects            \\
                       \cmidrule(rl){3-7}
                     & \multirow{2}{*}{\crossfire v1.4.0} & \pinned(1PT), 1vCPU each & \novuln    & \novuln             & \protects  & \protects            \\
                     &                                    & \pinned(2PT), 1vCPU each & \novuln    & \novuln             & \noprotect & \protects            \\
\midrule
\multirow{5}{*}{RSB} & \multirow{1}{*}{\host}             & any                      & \protects  & \protects           & \protects  & \protects            \\
                     & \multirow{1}{*}{\fire v1.0.0}      & any                      & \protects  & \protects           & \protects  & \protects            \\
                     & \multirow{1}{*}{\fire v1.4.0}      & any                      & \protects  & \protects           & \protects  & \protects            \\
                     & \multirow{1}{*}{\crossfire v1.0.0} & any                      & \novuln    & \novuln             & \protects  & \protects            \\
                     & \multirow{1}{*}{\crossfire v1.4.0} & any                      & \novuln    & \novuln             & \protects  & \protects            \\
\midrule
\multirow{3}{*}{STL} & \multirow{1}{*}{\host}             & any                      & \noprotect & \novuln             & \novuln    & \novuln              \\
                     & \multirow{1}{*}{\fire v1.0.0}      & any                      & \noprotect & \novuln             & \novuln    & \novuln              \\
                     & \multirow{1}{*}{\fire v1.4.0}      & any                      & \noprotect & \novuln             & \novuln    & \novuln              \\
\bottomrule
\multicolumn{7}{p{12cm}}{\protectskey} \\
\multicolumn{7}{l}{\host{} -- Experiments were run on the host system} \\
\multicolumn{7}{l}{\fire v1.x.0 -- Experiments were run inside a single Firecracker v1.x.0 VM} \\
\multicolumn{7}{l}{\crossfire v1.x.0 -- Experiments were run across two Firecracker v1.x.0 VMs} \\
\multicolumn{7}{l}{\pinned($X$PT) --  bare metal \acs{PoC} processes/Firecracker VM vCPUs were pinned to $X$ physical sibling threads.} \\
\multicolumn{7}{l}{$X$vCPU -- the Firecracker VM has $X$ virtual CPUs.} \\
\multicolumn{7}{l}{$^a$ the \acs{PoC} does not support running without being assigned to a core.}
\end{tabular}

\end{table*}

\begin{figure*}
    \begin{lstlisting}[gobble=8]
        Vulnerability Spec store bypass: Mitigation; Speculative Store Bypass disabled via prctl and seccomp
        Vulnerability Spectre v1:        Mitigation; usercopy/swapgs barriers and __user pointer sanitization
        Vulnerability Spectre v2:        Mitigation; Full generic retpoline, IBPB conditional, IBRS_FW, STIBP conditional, RSB filling
    \end{lstlisting}
    \caption{Spectre mitigations enabled in the host and guest kernel during the Spectre tests. This setup is recommended by \ac{AWS} for host production systems~\cite{amazon2023prodhostsetup}.}
    \label{lst:enabled-spectre-mitigations}
\end{figure*}

With \emph{\ac{AWS} recommended countermeasures}~\cite{amazon2023prodhostsetup} (the default for the Linux kernels in use, cf. \autoref{lst:enabled-spectre-mitigations}) enabled on the host system and inside Firecracker \acp{VM}, we see that Spectre-RSB is successfully mitigated both on the host and inside and across \acp{VM} (cf. \autoref{tab:spectre-pocs-protected}).
On the other hand, \emph{Spectre-STL, Spectre-BTB, and Spectre-PHT allowed information leakage} in particular situations.

The \acp{PoC} for Spectre-STL show leakage. However, the leakage only occurs within the same process and the same privilege level.
Since no cross-process variants are known, we didn't test the cross-VM scenario for Spectre-STL.
In our user-to-user threat model, Spectre-STL is not a possible attack vector, as no cross-process variants are known.
If two tenant workloads would be isolated by in-process isolation within the same \ac{VM}, Spectre-STL could still be a viable attack vector.
In the user-to-host model, Spectre-STL is mitigated by countermeasures that are included in current Linux kernels and enabled by default.

For Spectre-PHT, the kernel countermeasures include the sanitization of user-pointers and the utilization of barriers (\texttt{lfence}) on privilege level switches.
We therefore conclude that Spectre-PHT poses little to no threat to the host system.
However, these mitigations do not protect two hyperthreads from each other if they execute on the same physical core in parallel.
This is why all four Spectre-PHT mistraining variants are fully functional on the host system as well as inside Firecracker \acp{VM}.
As can be seen in \autoref{tab:spectre-pocs-protected}, this remains true \emph{even if \ac{SMT} is disabled}\footnote{This is simulated by pinning attacker and victim process to the same core (\pinned(1PT))}.
In fact, pinning both \acp{VM} to the same physical thread \emph{enables} the cross-process out-of-place version of Spectre-PHT whereas we did not observe leake in the \ac{SMT} case.
This makes Spectre-PHT a significant threat for user-to-user.

Spectre-BTB \acp{PoC} are partially functional when \ac{AWS} recommended countermeasures are enabled.
The original variant that mistrains the BTB in the same process and at the same address is fully functional while same-process out-of-place mistraining is successfully mitigated.
Also, all attempts to leak information across process boundaries via out-of-place mistraining did not show any leakage.
With cross-process in-place mistraining, however, we observed leakage.
On the host system, the leakage occurred independent of \ac{SMT}.
Inside a \ac{VM}, the leakage only occurred if all virtual CPU cores were assigned to a separate physical thread.
Across \acp{VM}, disabling \ac{SMT} removed the leakage.

Besides the countermeasures listed in \autoref{lst:enabled-spectre-mitigations}, the host kernel has Spectre countermeasures compiled into the \ac{VM} entry and exit point\footnote{\url{https://elixir.bootlin.com/linux/v5.10/source/arch/x86/kvm/vmx/vmenter.S\#L191}} to fully disable malicious guests from attacking the host kernel while the kernel handles a \ac{VM} exit.

In summary, we can say that the Linux default countermeasures--which are recommended by \ac{AWS} Firecracker--only partially mitigate Spectre.
Precisely, we show:
\begin{itemize}
    \item Spectre-PHT and Spectre-BTB can still leak information between tenants in the guest-to-guest scenario with the \ac{AWS} recommended countermeasures--which includes disabling \ac{SMT}--in place.
    \item The host kernel is likely sufficiently protected by the additional precautions that are compiled into the Linux kernel to shield \ac{VM} entries and exits. This, however, is orthogonal to security measures provided by Firecracker.
\end{itemize}

All leakage observed was independent of the Firecracker version in use.

\subsubsection{Evaluation}
We find that Firecracker does not add to the mitigations against Spectre but solely relies on general protection recommendations, which include mitigations provided by the host and guest kernels and optional microcode updates.
Even worse, the recommended countermeasures insufficiently protect serverless applications from leaking information to other tenants.
We therefore claim that Firecracker does \emph{not} achieve its isolation goal on a microarchitectural level, even though microarchitectural attacks are considered in-scope of the Firecracker threat model.

The alert reader might wonder why Spectre-BTB remains an issue with the STIBP countermeasure in place (cf. \autoref{lst:enabled-spectre-mitigations}) as this microcode patch was designed to stop the branch prediction from using prediction information that originates from another thread.
This also puzzled us for a while until recently Google published a security advisory\footnote{\url{https://github.com/google/security-research/security/advisories/GHSA-mj4w-6495-6crx}} that identifies a flaw in Linux 6.2 that kept disabling the STIBP mitigation when IBRS is enabled.
We verified that the code section that was identified as being responsible for the issue is also present in the Linux 5.10 source code.
Our assumption therefore is that the same problem identified by Google also occurs on our system.

\section{Conclusions}
Cloud technologies constantly shift to meet the needs of their customers.
At the same time, \acp{CSP} aim for maximizing efficiency and profit, which incentivizes serverless \acp{CSP} to over-commit available compute resources.
While this is reasonable from an economic perspective, the resulting system behavior can be disastrous in the context of microarchitectural attacks that exploit shared hardware resources.
In the past few years, the microarchitectural threat landscape changed frequently and rapidly.
There mitigations that work reasonably well to prevent many attacks, but they often lead to significant performance costs, which forces \acp{CSP} to find a tradeoff between economic value and security.
Furthermore, some microarchitectural attacks simply are not hindered by existing mitigations.
The \ac{CSP} customers have little control over the microarchitectural defenses deployed and must trust their providers to keep up with the pace of microarchitectural attack and mitigation development.
Defense-in-depth requires security at every level, from the microcode to \ac{VMM} to container.
Each system must be considered as a whole, as some protections at one system level may open a vulnerabilities at another.

We showed that default countermeasures as they are recommended for the Firecracker \ac{VMM} are insufficient to meet its isolation goals.
In fact, many of the tested attack vectors showed leakage while countermeasures where in place.
We identified the Medusa cache indexing/block write variant as an attack vector that only works across \acp{VM}, i.\,e. with additional isolation mechanisms in place.
Additionally, we showed that disabling \ac{SMT}--an expensive mitigation technique recommended and performed by \ac{AWS}--does not provide full protection against Medusa variants.
The aforementioned Medusa variant, and Spectre-PHT are still capable of leaking information between cloud tenants even if \ac{SMT} is disabled, as long as the attacker and target threads keep competing for hardware resources of the same physical CPU core.
Unfortunately this is inevitably the case in high-density serverless environments.
In the present, serverless \acp{CSP} must remain vigilant in keeping firmware up-to-date and employing all possible defenses against microarchitectural attacks.
Users must not only trust their \acp{CSP} of choice to keep their systems up-to-date and properly configured, but also be aware that some microarchitectural vulnerabilities, particularly certain Spectre variants, are still able to cross containment boundaries.
Furthermore, processor designs continue to evolve and speculative and out-of-order execution remain important factors in improving performance from generation to generation.
So, it is unlikely that we have seen the last of new microarchitectural vulnerabilities, as the recent wave of newly discovered attacks~\cite{trujillo2023inception,moghimi2023downfall,wilkner2023phantom} shows.


\begin{acks}
This work was supported by the \grantsponsor{dfg}{German Research Foundation (DFG)}{https://www.dfg.de/} under Grants No.~\grantnum{dfg}{439797619} and~\grantnum{dfg}{456967092}, by the \grantsponsor{bmbf}{German Federal Ministry for Education and Research (BMBF)}{https://www.bmbf.de} under Grants~\grantnum{bmbf}{SASVI} and~\grantnum{bmbf}{SILGENTAS}, by the \grantsponsor{nsf}{National Science Foundation (NSF)}{https://www.nsf.gov/} under Grant~\grantnum{nsf}{CNS-2026913}, and in part by a grant from the Qatar National Research Fund.
\end{acks}

\bibliographystyle{ACM-Reference-Format}
\bibliography{03_reference}


\begin{thebibliography}{55}


\ifx \showCODEN    \undefined \def \showCODEN     #1{\unskip}     \fi
\ifx \showDOI      \undefined \def \showDOI       #1{#1}\fi
\ifx \showISBNx    \undefined \def \showISBNx     #1{\unskip}     \fi
\ifx \showISBNxiii \undefined \def \showISBNxiii  #1{\unskip}     \fi
\ifx \showISSN     \undefined \def \showISSN      #1{\unskip}     \fi
\ifx \showLCCN     \undefined \def \showLCCN      #1{\unskip}     \fi
\ifx \shownote     \undefined \def \shownote      #1{#1}          \fi
\ifx \showarticletitle \undefined \def \showarticletitle #1{#1}   \fi
\ifx \showURL      \undefined \def \showURL       {\relax}        \fi
\providecommand\bibfield[2]{#2}
\providecommand\bibinfo[2]{#2}
\providecommand\natexlab[1]{#1}
\providecommand\showeprint[2][]{arXiv:#2}

\bibitem[\protect\citeauthoryear{Agache, Brooker, Iordache, Liguori,
  Neugebauer, Piwonka, and Popa}{Agache et~al\mbox{.}}{2020}]%
        {agache2020firecracker}
\bibfield{author}{\bibinfo{person}{Alexandru Agache}, \bibinfo{person}{Marc
  Brooker}, \bibinfo{person}{Alexandra Iordache}, \bibinfo{person}{Anthony
  Liguori}, \bibinfo{person}{Rolf Neugebauer}, \bibinfo{person}{Phil Piwonka},
  {and} \bibinfo{person}{Diana{-}Maria Popa}.} \bibinfo{year}{2020}\natexlab{}.
\newblock \showarticletitle{Firecracker: Lightweight Virtualization for
  Serverless Applications}. In \bibinfo{booktitle}{\emph{{NSDI}}}.
  \bibinfo{publisher}{{USENIX} Association}, \bibinfo{pages}{419--434}.
\newblock


\bibitem[\protect\citeauthoryear{Aldaya and Brumley}{Aldaya and
  Brumley}{2022}]%
        {aldaya2022hyperdegrade}
\bibfield{author}{\bibinfo{person}{Alejandro~Cabrera Aldaya} {and}
  \bibinfo{person}{Billy~Bob Brumley}.} \bibinfo{year}{2022}\natexlab{}.
\newblock \showarticletitle{HyperDegrade: From GHz to MHz Effective {CPU}
  Frequencies}. In \bibinfo{booktitle}{\emph{{USENIX} Security Symposium}}.
  \bibinfo{publisher}{{USENIX} Association}, \bibinfo{pages}{2801--2818}.
\newblock


\bibitem[\protect\citeauthoryear{Aldaya, Brumley, ul~Hassan, Garc{\'{\i}}a, and
  Tuveri}{Aldaya et~al\mbox{.}}{2019}]%
        {aldaya2019portcontention}
\bibfield{author}{\bibinfo{person}{Alejandro~Cabrera Aldaya},
  \bibinfo{person}{Billy~Bob Brumley}, \bibinfo{person}{Sohaib ul Hassan},
  \bibinfo{person}{Cesar~Pereida Garc{\'{\i}}a}, {and} \bibinfo{person}{Nicola
  Tuveri}.} \bibinfo{year}{2019}\natexlab{}.
\newblock \showarticletitle{Port Contention for Fun and Profit}. In
  \bibinfo{booktitle}{\emph{{IEEE} Symposium on Security and Privacy}}.
  \bibinfo{publisher}{{IEEE}}, \bibinfo{pages}{870--887}.
\newblock


\bibitem[\protect\citeauthoryear{{Amazon Web Services}}{{Amazon Web
  Services}}{2023a}]%
        {amazon2023fargate}
\bibfield{author}{\bibinfo{person}{{Amazon Web Services}}.}
  \bibinfo{year}{2023}\natexlab{a}.
\newblock \bibinfo{title}{{AWS} Fargate}.
\newblock
\newblock
\urldef\tempurl%
\url{https://docs.aws.amazon.com/eks/latest/userguide/fargate.html}
\showURL{%
\tempurl}
\newblock
\shownote{accessed: Aug 17, 2023.}


\bibitem[\protect\citeauthoryear{{Amazon Web Services}}{{Amazon Web
  Services}}{2023b}]%
        {amazon2023lambda}
\bibfield{author}{\bibinfo{person}{{Amazon Web Services}}.}
  \bibinfo{year}{2023}\natexlab{b}.
\newblock \bibinfo{title}{{AWS} Lambda Features}.
\newblock
\newblock
\urldef\tempurl%
\url{https://aws.amazon.com/lambda/features/}
\showURL{%
\tempurl}
\newblock
\shownote{accessed: Aug 17, 2023.}


\bibitem[\protect\citeauthoryear{{Amazon Web Services}}{{Amazon Web
  Services}}{2023c}]%
        {amazon2023firecrackerdesign}
\bibfield{author}{\bibinfo{person}{{Amazon Web Services}}.}
  \bibinfo{year}{2023}\natexlab{c}.
\newblock \bibinfo{title}{Firecracker Design}.
\newblock
  \bibinfo{howpublished}{\url{https://github.com/firecracker-microvm/firecracker/blob/9c51dc6852d68d0f6982a4017a63645fa75460c0/docs/design.md}}.
\newblock


\bibitem[\protect\citeauthoryear{{Amazon Web Services}}{{Amazon Web
  Services}}{2023d}]%
        {amazon2023jailer}
\bibfield{author}{\bibinfo{person}{{Amazon Web Services}}.}
  \bibinfo{year}{2023}\natexlab{d}.
\newblock \bibinfo{title}{The Firecracker Jailer}.
\newblock
  \bibinfo{howpublished}{\url{https://github.com/firecracker-microvm/firecracker/blob/main/docs/jailer.md}}.
\newblock
\newblock
\shownote{accessed: August 14, 2023.}


\bibitem[\protect\citeauthoryear{{Amazon Web Services}}{{Amazon Web
  Services}}{2023e}]%
        {amazon2023prodhostsetup}
\bibfield{author}{\bibinfo{person}{{Amazon Web Services}}.}
  \bibinfo{year}{2023}\natexlab{e}.
\newblock \bibinfo{title}{Production Host Setup Recommendations}.
\newblock
  \bibinfo{howpublished}{\url{https://github.com/firecracker-microvm/firecracker/blob/9ddeaf322a74c20cfb6b5af745112c95b7cecb75/docs/prod-host-setup.md}}.
\newblock
\newblock
\shownote{accessed: May 22, 2023.}


\bibitem[\protect\citeauthoryear{Balasubramanian, Baranowski, Burtsev, Panda,
  Rakamaric, and Ryzhyk}{Balasubramanian et~al\mbox{.}}{2017}]%
        {balasubramanian2017system}
\bibfield{author}{\bibinfo{person}{Abhiram Balasubramanian},
  \bibinfo{person}{Marek~S. Baranowski}, \bibinfo{person}{Anton Burtsev},
  \bibinfo{person}{Aurojit Panda}, \bibinfo{person}{Zvonimir Rakamaric}, {and}
  \bibinfo{person}{Leonid Ryzhyk}.} \bibinfo{year}{2017}\natexlab{}.
\newblock \showarticletitle{System Programming in Rust: Beyond Safety}. In
  \bibinfo{booktitle}{\emph{HotOS}}. \bibinfo{publisher}{{ACM}},
  \bibinfo{pages}{156--161}.
\newblock


\bibitem[\protect\citeauthoryear{Barberis, Frigo, Muench, Bos, and
  Giuffrida}{Barberis et~al\mbox{.}}{2022}]%
        {barberis2022branchhistoryinjection}
\bibfield{author}{\bibinfo{person}{Enrico Barberis}, \bibinfo{person}{Pietro
  Frigo}, \bibinfo{person}{Marius Muench}, \bibinfo{person}{Herbert Bos}, {and}
  \bibinfo{person}{Cristiano Giuffrida}.} \bibinfo{year}{2022}\natexlab{}.
\newblock \showarticletitle{Branch History Injection: On the Effectiveness of
  Hardware Mitigations Against Cross-Privilege Spectre-v2 Attacks}. In
  \bibinfo{booktitle}{\emph{{USENIX} Security Symposium}}.
  \bibinfo{publisher}{{USENIX} Association}, \bibinfo{pages}{971--988}.
\newblock


\bibitem[\protect\citeauthoryear{Bhattacharyya, Sandulescu, Neugschwandtner,
  Sorniotti, Falsafi, Payer, and Kurmus}{Bhattacharyya et~al\mbox{.}}{2019}]%
        {bhattacharyya2019smotherspectre}
\bibfield{author}{\bibinfo{person}{Atri Bhattacharyya},
  \bibinfo{person}{Alexandra Sandulescu}, \bibinfo{person}{Matthias
  Neugschwandtner}, \bibinfo{person}{Alessandro Sorniotti},
  \bibinfo{person}{Babak Falsafi}, \bibinfo{person}{Mathias Payer}, {and}
  \bibinfo{person}{Anil Kurmus}.} \bibinfo{year}{2019}\natexlab{}.
\newblock \showarticletitle{SMoTherSpectre: Exploiting Speculative Execution
  through Port Contention}. In \bibinfo{booktitle}{\emph{{CCS}}}.
  \bibinfo{publisher}{{ACM}}, \bibinfo{pages}{785--800}.
\newblock


\bibitem[\protect\citeauthoryear{Bulck, Moghimi, Schwarz, Lipp, Minkin, Genkin,
  Yarom, Sunar, Gruss, and Piessens}{Bulck et~al\mbox{.}}{2020}]%
        {vanbulk2020lvi}
\bibfield{author}{\bibinfo{person}{Jo~Van Bulck}, \bibinfo{person}{Daniel
  Moghimi}, \bibinfo{person}{Michael Schwarz}, \bibinfo{person}{Moritz Lipp},
  \bibinfo{person}{Marina Minkin}, \bibinfo{person}{Daniel Genkin},
  \bibinfo{person}{Yuval Yarom}, \bibinfo{person}{Berk Sunar},
  \bibinfo{person}{Daniel Gruss}, {and} \bibinfo{person}{Frank Piessens}.}
  \bibinfo{year}{2020}\natexlab{}.
\newblock \showarticletitle{{LVI:} Hijacking Transient Execution through
  Microarchitectural Load Value Injection}. In \bibinfo{booktitle}{\emph{{IEEE}
  Symposium on Security and Privacy}}. \bibinfo{publisher}{{IEEE}},
  \bibinfo{pages}{54--72}.
\newblock


\bibitem[\protect\citeauthoryear{Canella, Bulck, Schwarz, Lipp, von Berg,
  Ortner, Piessens, Evtyushkin, and Gruss}{Canella et~al\mbox{.}}{2019a}]%
        {canela2019systematicevaluation}
\bibfield{author}{\bibinfo{person}{Claudio Canella}, \bibinfo{person}{Jo~Van
  Bulck}, \bibinfo{person}{Michael Schwarz}, \bibinfo{person}{Moritz Lipp},
  \bibinfo{person}{Benjamin von Berg}, \bibinfo{person}{Philipp Ortner},
  \bibinfo{person}{Frank Piessens}, \bibinfo{person}{Dmitry Evtyushkin}, {and}
  \bibinfo{person}{Daniel Gruss}.} \bibinfo{year}{2019}\natexlab{a}.
\newblock \showarticletitle{A Systematic Evaluation of Transient Execution
  Attacks and Defenses}. In \bibinfo{booktitle}{\emph{{USENIX} Security
  Symposium}}. \bibinfo{publisher}{{USENIX} Association},
  \bibinfo{pages}{249--266}.
\newblock


\bibitem[\protect\citeauthoryear{Canella, Genkin, Giner, Gruss, Lipp, Minkin,
  Moghimi, Piessens, Schwarz, Sunar, Bulck, and Yarom}{Canella
  et~al\mbox{.}}{2019b}]%
        {canella2019fallout}
\bibfield{author}{\bibinfo{person}{Claudio Canella}, \bibinfo{person}{Daniel
  Genkin}, \bibinfo{person}{Lukas Giner}, \bibinfo{person}{Daniel Gruss},
  \bibinfo{person}{Moritz Lipp}, \bibinfo{person}{Marina Minkin},
  \bibinfo{person}{Daniel Moghimi}, \bibinfo{person}{Frank Piessens},
  \bibinfo{person}{Michael Schwarz}, \bibinfo{person}{Berk Sunar},
  \bibinfo{person}{Jo~Van Bulck}, {and} \bibinfo{person}{Yuval Yarom}.}
  \bibinfo{year}{2019}\natexlab{b}.
\newblock \showarticletitle{Fallout: Leaking Data on Meltdown-resistant CPUs}.
  In \bibinfo{booktitle}{\emph{{CCS}}}. \bibinfo{publisher}{{ACM}},
  \bibinfo{pages}{769--784}.
\newblock


\bibitem[\protect\citeauthoryear{Canella, {Van Bulck}, Schwarz, Gruss, Easdon,
  and Jha}{Canella et~al\mbox{.}}{2019c}]%
        {canela2019transientfail-code}
\bibfield{author}{\bibinfo{person}{Claudio Canella}, \bibinfo{person}{Jo {Van
  Bulck}}, \bibinfo{person}{Michael Schwarz}, \bibinfo{person}{Daniel Gruss},
  \bibinfo{person}{Catherine Easdon}, {and} \bibinfo{person}{Saagar Jha}.}
  \bibinfo{year}{2019}\natexlab{c}.
\newblock \bibinfo{title}{Transient Fail [Source Code]}.
\newblock
\newblock
\urldef\tempurl%
\url{https://github.com/IAIK/transientfail}
\showURL{%
\tempurl}


\bibitem[\protect\citeauthoryear{Chen, Chen, Xiao, Zhang, Lin, and Lai}{Chen
  et~al\mbox{.}}{2019}]%
        {chen2019sgxpectre}
\bibfield{author}{\bibinfo{person}{Guoxing Chen}, \bibinfo{person}{Sanchuan
  Chen}, \bibinfo{person}{Yuan Xiao}, \bibinfo{person}{Yinqian Zhang},
  \bibinfo{person}{Zhiqiang Lin}, {and} \bibinfo{person}{Ten{-}Hwang Lai}.}
  \bibinfo{year}{2019}\natexlab{}.
\newblock \showarticletitle{SgxPectre: Stealing Intel Secrets from {SGX}
  Enclaves Via Speculative Execution}. In
  \bibinfo{booktitle}{\emph{EuroS{\&}P}}. \bibinfo{publisher}{{IEEE}},
  \bibinfo{pages}{142--157}.
\newblock


\bibitem[\protect\citeauthoryear{Dole\u{z}elov{\'{a}}, Navr{\'{a}}til,
  Major\u{s}inov{\'{a}}, Ondrejka, Silas, Prpi\u{c}, and
  Landmann}{Dole\u{z}elov{\'{a}} et~al\mbox{.}}{2020}]%
        {redhat2020resource}
\bibfield{author}{\bibinfo{person}{Marie Dole\u{z}elov{\'{a}}},
  \bibinfo{person}{Milan Navr{\'{a}}til}, \bibinfo{person}{Eva
  Major\u{s}inov{\'{a}}}, \bibinfo{person}{Peter Ondrejka},
  \bibinfo{person}{Douglas Silas}, \bibinfo{person}{Martin Prpi\u{c}}, {and}
  \bibinfo{person}{R{\"{u}}diger Landmann}.} \bibinfo{year}{2020}\natexlab{}.
\newblock \bibinfo{booktitle}{\emph{Red Hat Enterprise Linux 7 Resource
  Management Guide--Using cgroups to manage system resources on RHEL}}.
\newblock Red Hat, Inc.
\newblock
\urldef\tempurl%
\url{https://access.redhat.com/documentation/en-us/red_hat_enterprise_linux/7/pdf/resource_management_guide/red_hat_enterprise_linux-7-resource_management_guide-en-us.pdf}
\showURL{%
\tempurl}
\newblock
\shownote{accessed: Aug 17, 2023.}


\bibitem[\protect\citeauthoryear{Fustos, Bechtel, and Yun}{Fustos
  et~al\mbox{.}}{2020}]%
        {fustos2020spectrerewind}
\bibfield{author}{\bibinfo{person}{Jacob Fustos},
  \bibinfo{person}{Michael~Garrett Bechtel}, {and} \bibinfo{person}{Heechul
  Yun}.} \bibinfo{year}{2020}\natexlab{}.
\newblock \showarticletitle{SpectreRewind: Leaking Secrets to Past
  Instructions}. In \bibinfo{booktitle}{\emph{ASHES@CCS}}.
  \bibinfo{publisher}{{ACM}}, \bibinfo{pages}{117--126}.
\newblock


\bibitem[\protect\citeauthoryear{Gruss, Lipp, Schwarz, Fellner, Maurice, and
  Mangard}{Gruss et~al\mbox{.}}{2017}]%
        {gruss2019kaslr}
\bibfield{author}{\bibinfo{person}{Daniel Gruss}, \bibinfo{person}{Moritz
  Lipp}, \bibinfo{person}{Michael Schwarz}, \bibinfo{person}{Richard Fellner},
  \bibinfo{person}{Cl{\'{e}}mentine Maurice}, {and} \bibinfo{person}{Stefan
  Mangard}.} \bibinfo{year}{2017}\natexlab{}.
\newblock \showarticletitle{{KASLR} is Dead: Long Live {KASLR}}. In
  \bibinfo{booktitle}{\emph{ESSoS}} \emph{(\bibinfo{series}{Lecture Notes in
  Computer Science}, Vol.~\bibinfo{volume}{10379})}.
  \bibinfo{publisher}{Springer}, \bibinfo{pages}{161--176}.
\newblock


\bibitem[\protect\citeauthoryear{Gupta}{Gupta}{2020}]%
        {linux2020taa}
\bibfield{author}{\bibinfo{person}{Pawan Gupta}.}
  \bibinfo{year}{2020}\natexlab{}.
\newblock \bibinfo{booktitle}{\emph{{TAA} - {TSX} Asynchronous Abort}}.
\newblock The Linux Kernel Organization.
\newblock
\urldef\tempurl%
\url{https://www.kernel.org/doc/html/latest/admin-guide/hw-vuln/tsx_async_abort.html}
\showURL{%
\tempurl}
\newblock
\shownote{accessed: Aug 17, 2023.}


\bibitem[\protect\citeauthoryear{Hicks}{Hicks}{2019}]%
        {linux2019mds}
\bibfield{author}{\bibinfo{person}{Tyler Hicks}.}
  \bibinfo{year}{2019}\natexlab{}.
\newblock \bibinfo{booktitle}{\emph{{MDS} - Microarchitectural Data Sampling}}.
\newblock The Linux Kernel Organization.
\newblock
\urldef\tempurl%
\url{https://www.kernel.org/doc/html/latest/admin-guide/hw-vuln/mds.html}
\showURL{%
\tempurl}
\newblock
\shownote{accessed: Aug 17, 2023.}


\bibitem[\protect\citeauthoryear{Horn}{Horn}{2018}]%
        {horn2018sectrev4}
\bibfield{author}{\bibinfo{person}{Jann Horn}.}
  \bibinfo{year}{2018}\natexlab{}.
\newblock \bibinfo{title}{Speculative execution, variant 4: speculative store
  bypass}.
\newblock
\newblock
\urldef\tempurl%
\url{https://bugs.chromium.org/p/project-zero/issues/detail?id=1528}
\showURL{%
\tempurl}
\newblock
\shownote{accessed: Aug 17, 2023.}


\bibitem[\protect\citeauthoryear{Intel}{Intel}{2018}]%
        {intel2018spectremitigation}
\bibfield{author}{\bibinfo{person}{Intel}.} \bibinfo{year}{2018}\natexlab{}.
\newblock \bibinfo{title}{Speculative Execution Side Channel Mitigations}.
\newblock
  \bibinfo{howpublished}{\url{https://www.intel.com/content/dam/develop/external/us/en/documents/336996-speculative-execution-side-channel-mitigations.pdf}}.
\newblock
\newblock
\shownote{rev. 3.0 accessed: Mar 22, 2023.}


\bibitem[\protect\citeauthoryear{Intel}{Intel}{2019a}]%
        {intel2019taa}
\bibfield{author}{\bibinfo{person}{Intel}.} \bibinfo{year}{2019}\natexlab{a}.
\newblock \bibinfo{booktitle}{\emph{{Intel} Transactional Synchronization
  Extensions ({Intel TSX}) Asynchronous Abort}}.
\newblock \bibinfo{type}{{T}echnical {R}eport}. \bibinfo{institution}{Intel
  Corp.}
\newblock
\urldef\tempurl%
\url{https://www.intel.com/content/www/us/en/developer/articles/technical/software-security-guidance/technical-documentation/intel-tsx-asynchronous-abort.html}
\showURL{%
\tempurl}
\newblock
\shownote{accessed: Aug 17, 2023.}


\bibitem[\protect\citeauthoryear{Intel}{Intel}{2019b}]%
        {intel2019mds}
\bibfield{author}{\bibinfo{person}{Intel}.} \bibinfo{year}{2019}\natexlab{b}.
\newblock \bibinfo{booktitle}{\emph{Microarchitectural Data Sampling}}.
\newblock \bibinfo{type}{{T}echnical {R}eport}. \bibinfo{institution}{Intel
  Corp.}
\newblock
\urldef\tempurl%
\url{https://www.intel.com/content/www/us/en/developer/articles/technical/software-security-guidance/technical-documentation/intel-analysis-microarchitectural-data-sampling.html}
\showURL{%
\tempurl}
\newblock
\shownote{ver. 3.0, accessed: Aug 17, 2023.}


\bibitem[\protect\citeauthoryear{Intel}{Intel}{2020}]%
        {intel2020vrs}
\bibfield{author}{\bibinfo{person}{Intel}.} \bibinfo{year}{2020}\natexlab{}.
\newblock \bibinfo{booktitle}{\emph{Vector Register Sampling}}.
\newblock \bibinfo{type}{{T}echnical {R}eport}. \bibinfo{institution}{Intel
  Corp.}
\newblock
\urldef\tempurl%
\url{https://www.intel.com/content/www/us/en/developer/articles/technical/software-security-guidance/advisory-guidance/vector-register-sampling.html}
\showURL{%
\tempurl}
\newblock
\shownote{accessed: Aug 17, 2023.}


\bibitem[\protect\citeauthoryear{Johannesmeyer, Koschel, Razavi, Bos, and
  Giuffrida}{Johannesmeyer et~al\mbox{.}}{2022}]%
        {johannesmeyer2022kasper}
\bibfield{author}{\bibinfo{person}{Brian Johannesmeyer}, \bibinfo{person}{Jakob
  Koschel}, \bibinfo{person}{Kaveh Razavi}, \bibinfo{person}{Herbert Bos},
  {and} \bibinfo{person}{Cristiano Giuffrida}.}
  \bibinfo{year}{2022}\natexlab{}.
\newblock \showarticletitle{Kasper: Scanning for Generalized Transient
  Execution Gadgets in the Linux Kernel}. In
  \bibinfo{booktitle}{\emph{{NDSS}}}. \bibinfo{publisher}{The Internet
  Society}.
\newblock


\bibitem[\protect\citeauthoryear{Kiriansky and Waldspurger}{Kiriansky and
  Waldspurger}{2018}]%
        {kiriansky2018speculativebufferoverflows}
\bibfield{author}{\bibinfo{person}{Vladimir Kiriansky} {and}
  \bibinfo{person}{Carl~A. Waldspurger}.} \bibinfo{year}{2018}\natexlab{}.
\newblock \showarticletitle{Speculative Buffer Overflows: Attacks and
  Defenses}.
\newblock \bibinfo{journal}{\emph{CoRR}}  \bibinfo{volume}{abs/1807.03757}
  (\bibinfo{year}{2018}).
\newblock


\bibitem[\protect\citeauthoryear{Kivity, Kamay, Laor, Lubin, and
  Liguori}{Kivity et~al\mbox{.}}{2007}]%
        {kivity2007kvm}
\bibfield{author}{\bibinfo{person}{Avi Kivity}, \bibinfo{person}{Yaniv Kamay},
  \bibinfo{person}{Dor Laor}, \bibinfo{person}{Uri Lubin}, {and}
  \bibinfo{person}{Anthony Liguori}.} \bibinfo{year}{2007}\natexlab{}.
\newblock \showarticletitle{kvm: the Linux Virtual Machine Monitor}. In
  \bibinfo{booktitle}{\emph{Linux Symposium}}, Vol.~\bibinfo{volume}{1}.
  \bibinfo{publisher}{{kernel.org}}, \bibinfo{pages}{225--230}.
\newblock


\bibitem[\protect\citeauthoryear{Kocher, Horn, Fogh, Genkin, Gruss, Haas,
  Hamburg, Lipp, Mangard, Prescher, Schwarz, and Yarom}{Kocher
  et~al\mbox{.}}{2019}]%
        {kocher2019spectre}
\bibfield{author}{\bibinfo{person}{Paul Kocher}, \bibinfo{person}{Jann Horn},
  \bibinfo{person}{Anders Fogh}, \bibinfo{person}{Daniel Genkin},
  \bibinfo{person}{Daniel Gruss}, \bibinfo{person}{Werner Haas},
  \bibinfo{person}{Mike Hamburg}, \bibinfo{person}{Moritz Lipp},
  \bibinfo{person}{Stefan Mangard}, \bibinfo{person}{Thomas Prescher},
  \bibinfo{person}{Michael Schwarz}, {and} \bibinfo{person}{Yuval Yarom}.}
  \bibinfo{year}{2019}\natexlab{}.
\newblock \showarticletitle{Spectre Attacks: Exploiting Speculative Execution}.
  In \bibinfo{booktitle}{\emph{{IEEE} Symposium on Security and Privacy}}.
  \bibinfo{publisher}{{IEEE}}, \bibinfo{pages}{1--19}.
\newblock


\bibitem[\protect\citeauthoryear{Koruyeh, Khasawneh, Song, and
  Abu{-}Ghazaleh}{Koruyeh et~al\mbox{.}}{2018}]%
        {koruyeh2018spectreReturns}
\bibfield{author}{\bibinfo{person}{Esmaeil~Mohammadian Koruyeh},
  \bibinfo{person}{Khaled~N. Khasawneh}, \bibinfo{person}{Chengyu Song}, {and}
  \bibinfo{person}{Nael~B. Abu{-}Ghazaleh}.} \bibinfo{year}{2018}\natexlab{}.
\newblock \showarticletitle{Spectre Returns! Speculation Attacks using the
  Return Stack Buffer}. In \bibinfo{booktitle}{\emph{{WOOT} @ {USENIX} Security
  Symposium}}. \bibinfo{publisher}{{USENIX} Association}.
\newblock


\bibitem[\protect\citeauthoryear{Lipp, Schwarz, Gruss, Prescher, Haas, Fogh,
  Horn, Mangard, Kocher, Genkin, Yarom, and Hamburg}{Lipp
  et~al\mbox{.}}{2018}]%
        {lipp2018meltdown}
\bibfield{author}{\bibinfo{person}{Moritz Lipp}, \bibinfo{person}{Michael
  Schwarz}, \bibinfo{person}{Daniel Gruss}, \bibinfo{person}{Thomas Prescher},
  \bibinfo{person}{Werner Haas}, \bibinfo{person}{Anders Fogh},
  \bibinfo{person}{Jann Horn}, \bibinfo{person}{Stefan Mangard},
  \bibinfo{person}{Paul Kocher}, \bibinfo{person}{Daniel Genkin},
  \bibinfo{person}{Yuval Yarom}, {and} \bibinfo{person}{Mike Hamburg}.}
  \bibinfo{year}{2018}\natexlab{}.
\newblock \showarticletitle{Meltdown: Reading Kernel Memory from User Space}.
  In \bibinfo{booktitle}{\emph{{USENIX} Security Symposium}}.
  \bibinfo{publisher}{{USENIX} Association}, \bibinfo{pages}{973--990}.
\newblock


\bibitem[\protect\citeauthoryear{Maisuradze and Rossow}{Maisuradze and
  Rossow}{2018}]%
        {maisuradze2018ret2spec}
\bibfield{author}{\bibinfo{person}{Giorgi Maisuradze} {and}
  \bibinfo{person}{Christian Rossow}.} \bibinfo{year}{2018}\natexlab{}.
\newblock \showarticletitle{ret2spec: Speculative Execution Using Return Stack
  Buffers}. In \bibinfo{booktitle}{\emph{{CCS}}}. \bibinfo{publisher}{{ACM}},
  \bibinfo{pages}{2109--2122}.
\newblock


\bibitem[\protect\citeauthoryear{Marr, Binns, Hill, Hinton, Koufaty, Miller,
  and Upton}{Marr et~al\mbox{.}}{2002}]%
        {intel2002hyperthreading}
\bibfield{author}{\bibinfo{person}{Debora~T. Marr}, \bibinfo{person}{Frank
  Binns}, \bibinfo{person}{David~L. Hill}, \bibinfo{person}{Glenn Hinton},
  \bibinfo{person}{David~A. Koufaty}, \bibinfo{person}{J.~Alan Miller}, {and}
  \bibinfo{person}{Michael Upton}.} \bibinfo{year}{2002}\natexlab{}.
\newblock \showarticletitle{Hyper-Threading Technology Architecture and
  Microarchitecture}.
\newblock \bibinfo{journal}{\emph{Intel Technology Journal}}
  \bibinfo{volume}{6}, \bibinfo{number}{1} (\bibinfo{year}{2002}),
  \bibinfo{pages}{4--15}.
\newblock


\bibitem[\protect\citeauthoryear{Moghimi}{Moghimi}{2020}]%
        {moghimi2020medusa-code}
\bibfield{author}{\bibinfo{person}{Daniel Moghimi}.}
  \bibinfo{year}{2020}\natexlab{}.
\newblock \bibinfo{title}{Medusa Code Repository [Source Code]}.
\newblock
\newblock
\urldef\tempurl%
\url{https://github.com/flowyroll/medusa}
\showURL{%
\tempurl}


\bibitem[\protect\citeauthoryear{Moghimi}{Moghimi}{2023}]%
        {moghimi2023downfall}
\bibfield{author}{\bibinfo{person}{Daniel Moghimi}.}
  \bibinfo{year}{2023}\natexlab{}.
\newblock \showarticletitle{Downfall: Exploiting Speculative Data Gathering}.
  In \bibinfo{booktitle}{\emph{{USENIX} Security Symposium}}.
  \bibinfo{publisher}{{USENIX} Association}, \bibinfo{pages}{7179--7193}.
\newblock


\bibitem[\protect\citeauthoryear{Moghimi, Lipp, Sunar, and Schwarz}{Moghimi
  et~al\mbox{.}}{2020}]%
        {moghimi2020medusa}
\bibfield{author}{\bibinfo{person}{Daniel Moghimi}, \bibinfo{person}{Moritz
  Lipp}, \bibinfo{person}{Berk Sunar}, {and} \bibinfo{person}{Michael
  Schwarz}.} \bibinfo{year}{2020}\natexlab{}.
\newblock \showarticletitle{Medusa: Microarchitectural Data Leakage via
  Automated Attack Synthesis}. In \bibinfo{booktitle}{\emph{{USENIX} Security
  Symposium}}. \bibinfo{publisher}{{USENIX} Association},
  \bibinfo{pages}{1427--1444}.
\newblock


\bibitem[\protect\citeauthoryear{Narayan, Disselkoen, Moghimi, Cauligi,
  Johnson, Gang, Vahldiek{-}Oberwagner, Sahita, Shacham, Tullsen, and
  Stefan}{Narayan et~al\mbox{.}}{2021}]%
        {narayan2021swivel}
\bibfield{author}{\bibinfo{person}{Shravan Narayan}, \bibinfo{person}{Craig
  Disselkoen}, \bibinfo{person}{Daniel Moghimi}, \bibinfo{person}{Sunjay
  Cauligi}, \bibinfo{person}{Evan Johnson}, \bibinfo{person}{Zhao Gang},
  \bibinfo{person}{Anjo Vahldiek{-}Oberwagner}, \bibinfo{person}{Ravi Sahita},
  \bibinfo{person}{Hovav Shacham}, \bibinfo{person}{Dean~M. Tullsen}, {and}
  \bibinfo{person}{Deian Stefan}.} \bibinfo{year}{2021}\natexlab{}.
\newblock \showarticletitle{Swivel: Hardening WebAssembly against Spectre}. In
  \bibinfo{booktitle}{\emph{{USENIX} Security Symposium}}.
  \bibinfo{publisher}{{USENIX} Association}, \bibinfo{pages}{1433--1450}.
\newblock


\bibitem[\protect\citeauthoryear{Osvik, Shamir, and Tromer}{Osvik
  et~al\mbox{.}}{2006}]%
        {osvik2006primeprobe}
\bibfield{author}{\bibinfo{person}{Dag~Arne Osvik}, \bibinfo{person}{Adi
  Shamir}, {and} \bibinfo{person}{Eran Tromer}.}
  \bibinfo{year}{2006}\natexlab{}.
\newblock \showarticletitle{Cache Attacks and Countermeasures: The Case of
  {AES}}. In \bibinfo{booktitle}{\emph{{CT-RSA}}}
  \emph{(\bibinfo{series}{Lecture Notes in Computer Science},
  Vol.~\bibinfo{volume}{3860})}. \bibinfo{publisher}{Springer},
  \bibinfo{pages}{1--20}.
\newblock


\bibitem[\protect\citeauthoryear{Purnal, Turan, and Verbauwhede}{Purnal
  et~al\mbox{.}}{2021}]%
        {purnal2021primescope}
\bibfield{author}{\bibinfo{person}{Antoon Purnal}, \bibinfo{person}{Furkan
  Turan}, {and} \bibinfo{person}{Ingrid Verbauwhede}.}
  \bibinfo{year}{2021}\natexlab{}.
\newblock \showarticletitle{Prime+Scope: Overcoming the Observer Effect for
  High-Precision Cache Contention Attacks}. In
  \bibinfo{booktitle}{\emph{{CCS}}}. \bibinfo{publisher}{{ACM}},
  \bibinfo{pages}{2906--2920}.
\newblock


\bibitem[\protect\citeauthoryear{{Qumranet Inc}}{{Qumranet Inc}}{2006}]%
        {qumranet2006kvm}
\bibfield{author}{\bibinfo{person}{{Qumranet Inc}}.}
  \bibinfo{year}{2006}\natexlab{}.
\newblock \bibinfo{booktitle}{\emph{{KVM}: Kernel-based Virtualization Driver,
  White Paper}}.
\newblock \bibinfo{type}{{T}echnical {R}eport}. \bibinfo{institution}{Qumranet
  Inc}.
\newblock
\urldef\tempurl%
\url{https://docs.huihoo.com/kvm/kvm-white-paper.pdf}
\showURL{%
\tempurl}
\newblock
\shownote{accessed: Aug 17, 2023.}


\bibitem[\protect\citeauthoryear{Ragab, Barberis, Bos, and Giuffrida}{Ragab
  et~al\mbox{.}}{2021}]%
        {ragab2021machineclear}
\bibfield{author}{\bibinfo{person}{Hany Ragab}, \bibinfo{person}{Enrico
  Barberis}, \bibinfo{person}{Herbert Bos}, {and} \bibinfo{person}{Cristiano
  Giuffrida}.} \bibinfo{year}{2021}\natexlab{}.
\newblock \showarticletitle{Rage Against the Machine Clear: {A} Systematic
  Analysis of Machine Clears and Their Implications for Transient Execution
  Attacks}. In \bibinfo{booktitle}{\emph{{USENIX} Security Symposium}}.
  \bibinfo{publisher}{{USENIX} Association}, \bibinfo{pages}{1451--1468}.
\newblock


\bibitem[\protect\citeauthoryear{Rokicki, Maurice, Botvinnik, and Oren}{Rokicki
  et~al\mbox{.}}{2022b}]%
        {rokicki2022portableportcontention}
\bibfield{author}{\bibinfo{person}{Thomas Rokicki},
  \bibinfo{person}{Cl{\'{e}}mentine Maurice}, \bibinfo{person}{Marina
  Botvinnik}, {and} \bibinfo{person}{Yossi Oren}.}
  \bibinfo{year}{2022}\natexlab{b}.
\newblock \showarticletitle{Port Contention Goes Portable: Port Contention Side
  Channels in Web Browsers}. In \bibinfo{booktitle}{\emph{AsiaCCS}}.
  \bibinfo{publisher}{{ACM}}, \bibinfo{pages}{1182--1194}.
\newblock


\bibitem[\protect\citeauthoryear{Rokicki, Maurice, and Schwarz}{Rokicki
  et~al\mbox{.}}{2022a}]%
        {rokicki2022portcontentionwosmt}
\bibfield{author}{\bibinfo{person}{Thomas Rokicki},
  \bibinfo{person}{Cl{\'{e}}mentine Maurice}, {and} \bibinfo{person}{Michael
  Schwarz}.} \bibinfo{year}{2022}\natexlab{a}.
\newblock \showarticletitle{{CPU} Port Contention Without {SMT}}. In
  \bibinfo{booktitle}{\emph{{ESORICS} {(3)}}} \emph{(\bibinfo{series}{Lecture
  Notes in Computer Science}, Vol.~\bibinfo{volume}{13556})}.
  \bibinfo{publisher}{Springer}, \bibinfo{pages}{209--228}.
\newblock


\bibitem[\protect\citeauthoryear{Schrammel, Weiser, Steinegger, Schwarzl,
  Schwarz, Mangard, and Gruss}{Schrammel et~al\mbox{.}}{2020}]%
        {schrammel2020donky}
\bibfield{author}{\bibinfo{person}{David Schrammel}, \bibinfo{person}{Samuel
  Weiser}, \bibinfo{person}{Stefan Steinegger}, \bibinfo{person}{Martin
  Schwarzl}, \bibinfo{person}{Michael Schwarz}, \bibinfo{person}{Stefan
  Mangard}, {and} \bibinfo{person}{Daniel Gruss}.}
  \bibinfo{year}{2020}\natexlab{}.
\newblock \showarticletitle{Donky: Domain Keys - Efficient In-Process Isolation
  for {RISC-V} and x86}. In \bibinfo{booktitle}{\emph{{USENIX} Security
  Symposium}}. \bibinfo{publisher}{{USENIX} Association},
  \bibinfo{pages}{1677--1694}.
\newblock


\bibitem[\protect\citeauthoryear{Schwarz, Lipp, Moghimi, Bulck, Stecklina,
  Prescher, and Gruss}{Schwarz et~al\mbox{.}}{2019}]%
        {schwarz2019zombieload}
\bibfield{author}{\bibinfo{person}{Michael Schwarz}, \bibinfo{person}{Moritz
  Lipp}, \bibinfo{person}{Daniel Moghimi}, \bibinfo{person}{Jo~Van Bulck},
  \bibinfo{person}{Julian Stecklina}, \bibinfo{person}{Thomas Prescher}, {and}
  \bibinfo{person}{Daniel Gruss}.} \bibinfo{year}{2019}\natexlab{}.
\newblock \showarticletitle{ZombieLoad: Cross-Privilege-Boundary Data
  Sampling}. In \bibinfo{booktitle}{\emph{{CCS}}}. \bibinfo{publisher}{{ACM}},
  \bibinfo{pages}{753--768}.
\newblock


\bibitem[\protect\citeauthoryear{Trujillo, Wikner, and Razavi}{Trujillo
  et~al\mbox{.}}{2023}]%
        {trujillo2023inception}
\bibfield{author}{\bibinfo{person}{Daniël Trujillo}, \bibinfo{person}{Johannes
  Wikner}, {and} \bibinfo{person}{Kaveh Razavi}.}
  \bibinfo{year}{2023}\natexlab{}.
\newblock \showarticletitle{Inception: Exposing New Attack Surfaces with
  Training in Transient Execution}. In \bibinfo{booktitle}{\emph{{USENIX}
  Security Symposium}}. \bibinfo{publisher}{{USENIX} Association},
  \bibinfo{pages}{7303--7320}.
\newblock


\bibitem[\protect\citeauthoryear{Turner}{Turner}{2018}]%
        {google2018retpoline}
\bibfield{author}{\bibinfo{person}{Paul Turner}.}
  \bibinfo{year}{2018}\natexlab{}.
\newblock \bibinfo{title}{Retpoline: a software construct for preventing
  branch-target-injection}.
\newblock
  \bibinfo{howpublished}{\url{https://support.google.com/faqs/answer/7625886}}.
\newblock
\newblock
\shownote{accessed: Mar 22, 2023.}


\bibitem[\protect\citeauthoryear{Vahldiek{-}Oberwagner, Elnikety, Duarte,
  Sammler, Druschel, and Garg}{Vahldiek{-}Oberwagner et~al\mbox{.}}{2019}]%
        {vahldiek2019erim}
\bibfield{author}{\bibinfo{person}{Anjo Vahldiek{-}Oberwagner},
  \bibinfo{person}{Eslam Elnikety}, \bibinfo{person}{Nuno~O. Duarte},
  \bibinfo{person}{Michael Sammler}, \bibinfo{person}{Peter Druschel}, {and}
  \bibinfo{person}{Deepak Garg}.} \bibinfo{year}{2019}\natexlab{}.
\newblock \showarticletitle{{ERIM:} Secure, Efficient In-process Isolation with
  Protection Keys {(MPK)}}. In \bibinfo{booktitle}{\emph{{USENIX} Security
  Symposium}}. \bibinfo{publisher}{{USENIX} Association},
  \bibinfo{pages}{1221--1238}.
\newblock


\bibitem[\protect\citeauthoryear{van Schaik, Milburn, {\"{O}}sterlund, Frigo,
  Maisuradze, Razavi, Bos, and Giuffrida}{van Schaik et~al\mbox{.}}{2019}]%
        {vanschaik2019ridl}
\bibfield{author}{\bibinfo{person}{Stephan van Schaik}, \bibinfo{person}{Alyssa
  Milburn}, \bibinfo{person}{Sebastian {\"{O}}sterlund},
  \bibinfo{person}{Pietro Frigo}, \bibinfo{person}{Giorgi Maisuradze},
  \bibinfo{person}{Kaveh Razavi}, \bibinfo{person}{Herbert Bos}, {and}
  \bibinfo{person}{Cristiano Giuffrida}.} \bibinfo{year}{2019}\natexlab{}.
\newblock \showarticletitle{{RIDL:} Rogue In-Flight Data Load}. In
  \bibinfo{booktitle}{\emph{{IEEE} Symposium on Security and Privacy}}.
  \bibinfo{publisher}{{IEEE}}, \bibinfo{pages}{88--105}.
\newblock


\bibitem[\protect\citeauthoryear{van Schaik, Millburn, genBTC, Menzel, jun1x,
  Kitt, pit fr, {\"O}sterlund, and Giuffrida}{van Schaik et~al\mbox{.}}{2020}]%
        {vanschaik2020ridl-code}
\bibfield{author}{\bibinfo{person}{Stephan van Schaik}, \bibinfo{person}{Alyssa
  Millburn}, \bibinfo{person}{genBTC}, \bibinfo{person}{Paul Menzel},
  \bibinfo{person}{jun1x}, \bibinfo{person}{Stephen Kitt}, \bibinfo{person}{pit
  fr}, \bibinfo{person}{Sebastian {\"O}sterlund}, {and}
  \bibinfo{person}{Cristiano Giuffrida}.} \bibinfo{year}{2020}\natexlab{}.
\newblock \bibinfo{title}{RIDL [Source Code]}.
\newblock
\newblock
\urldef\tempurl%
\url{https://github.com/vusec/ridl}
\showURL{%
\tempurl}


\bibitem[\protect\citeauthoryear{Wikner and Razavi}{Wikner and Razavi}{2022}]%
        {wikner2022retbleed}
\bibfield{author}{\bibinfo{person}{Johannes Wikner} {and}
  \bibinfo{person}{Kaveh Razavi}.} \bibinfo{year}{2022}\natexlab{}.
\newblock \showarticletitle{{RETBLEED:} Arbitrary Speculative Code Execution
  with Return Instructions}. In \bibinfo{booktitle}{\emph{{USENIX} Security
  Symposium}}. \bibinfo{publisher}{{USENIX} Association},
  \bibinfo{pages}{3825--3842}.
\newblock


\bibitem[\protect\citeauthoryear{Wikner, Trujillo, and Razav}{Wikner
  et~al\mbox{.}}{2023}]%
        {wilkner2023phantom}
\bibfield{author}{\bibinfo{person}{Johannes Wikner}, \bibinfo{person}{Daniël
  Trujillo}, {and} \bibinfo{person}{Kaveh Razav}.}
  \bibinfo{year}{2023}\natexlab{}.
\newblock \showarticletitle{Phantom: Exploiting Decoder-detectable
  Mispredictions}. In \bibinfo{booktitle}{\emph{{MICRO} (to appear)}}.
  \bibinfo{publisher}{{IEEE}}.
\newblock


\bibitem[\protect\citeauthoryear{Yarom and Falkner}{Yarom and Falkner}{2014}]%
        {yarom2014flushreload}
\bibfield{author}{\bibinfo{person}{Yuval Yarom} {and} \bibinfo{person}{Katrina
  Falkner}.} \bibinfo{year}{2014}\natexlab{}.
\newblock \showarticletitle{{FLUSH+RELOAD:} {A} High Resolution, Low Noise,
  {L3} Cache Side-Channel Attack}. In \bibinfo{booktitle}{\emph{{USENIX}
  Security Symposium}}. \bibinfo{publisher}{{USENIX} Association},
  \bibinfo{pages}{719--732}.
\newblock


\bibitem[\protect\citeauthoryear{Young, Zhu, Caraza{-}Harter, Arpaci{-}Dusseau,
  and Arpaci{-}Dusseau}{Young et~al\mbox{.}}{2019}]%
        {young2019true}
\bibfield{author}{\bibinfo{person}{Ethan~G. Young}, \bibinfo{person}{Pengfei
  Zhu}, \bibinfo{person}{Tyler Caraza{-}Harter}, \bibinfo{person}{Andrea~C.
  Arpaci{-}Dusseau}, {and} \bibinfo{person}{Remzi~H. Arpaci{-}Dusseau}.}
  \bibinfo{year}{2019}\natexlab{}.
\newblock \showarticletitle{The True Cost of Containing: {A} gVisor Case
  Study}. In \bibinfo{booktitle}{\emph{HotCloud}}. \bibinfo{publisher}{{USENIX}
  Association}.
\newblock


\end{thebibliography}

\appendix

\end{document}